\documentclass[aps,pre,groupedaddress,showpacs,preprint]{revtex4-1}

\usepackage[dvips]{epsfig}% Include figure files
\usepackage{dcolumn}% Align table columns on decimal point
\usepackage{subfigure}%
\usepackage{bm}% bold math
\usepackage{amssymb}
\usepackage{amsmath}
\usepackage{color,soul}
\usepackage{cases}

\usepackage[latin1]{inputenc}

\usepackage{color,soul}  %%%%%%%%%%%%%%%%%%%%%
\setstcolor{red}         %%%%%%%%%%%%%%%%%%%%%
\newcommand{\be}{\begin{equation}}
\newcommand{\ee}{\end{equation}}
\newcommand{\ba}{\begin{eqnarray}}
\newcommand{\ea}{\end{eqnarray}}
\newcommand{\besu}{\begin{subequations}}
\newcommand{\esu}{\end{subequations}}
\newcommand{\nn}{\nonumber}

\newcommand{\ancho}{0.65}

\begin{document}
\title{Encounter-controlled coalescence and annihilation on a one-dimensional growing domain}
\author{F. Le Vot$^{1}$,  C. Escudero$^{2}$,  E. Abad$^{3}$, and S. B. Yuste$^{1}$}
\affiliation{
 $^{1}$ Departamento de F\'{\i}sica and Instituto de Computaci\'on Cient\'{\i}fica Avanzada (ICCAEx) \\
 Universidad de Extremadura, E-06071 Badajoz, Spain \\
$^{2}$
Departamento de Matem\'aticas \\
Universidad Aut\'onoma de Madrid \\
and Instituto de Ciencias Matem\'aticas \\
Consejo Superior de Investigaciones Cient\'{\i}ficas,\\
E-28049 Madrid, Spain \\
$^{3}$ Departamento de F\'{\i}sica Aplicada and Instituto de Computaci\'on Cient\'{\i}fica Avanzada (ICCAEx) \\ Centro Universitario de M\'erida \\ Universidad de Extremadura, E-06800 M\'erida, Spain \\
}

\begin{abstract}
The kinetics of encounter-controlled processes in growing domains is markedly different from that in a static domain. Here, we consider the specific example of diffusion limited coalescence and annihilation reactions in one-dimensional space. In the static case, such reactions are among the few systems amenable to exact solution, which can be obtained by means of a well-known method of intervals. In the case of a uniformly growing domain, we show that a double transformation in time and space allows one to extend this method to compute the main quantities characterizing the spatial and temporal behavior. We show that a sufficiently fast domain growth brings about drastic changes in the behavior. In this case, the reactions stop prematurely, as a result of which the survival probability of the reacting particles tends to a finite value at long times and their spatial distribution freezes before reaching the fully self-ordered state. We obtain exact results for the survival probability and for key properties characterizing the degree of self-ordering induced by the chemical reactions, i.e., the interparticle distribution function and the pair correlation function. These results are confirmed by numerical simulations.
\end{abstract}


\maketitle

\section{Introduction}
\label{intro}

In many instances, mass transport processes and chemical reactions are found to occur on different times scales. If the transport process facilitating the encounter of a pair of co-reactants is fast in comparison with the transformation process due to the reaction (as is for instance the case in a well-stirred system) spatial inhomogeneities are short-lived and local concentration fluctuations become negligible after an initial-condition-dependent transient regime. In this reaction-controlled regime, a mean-field description relying on classical rate equations provides a rather satisfactory, semiquantitative description of the underlying kinetics. In the present work, we consider the opposite limit where the typical reaction time is short with respect to the typical encounter time of a pair of diffusing co-reactants. That is, we consider the case of a diffusion-controlled reaction.

The theory of diffusion-controlled reactions was strongly influenced by the pioneering works of Smoluchowski, Collins and Kimball, and Noyes \cite{Rice1985}, and has ever since provided an extremely successful description of the kinetics of many systems of practical importance, such as fluorescence/luminiscence quenching, reactions of the solvated electron, proton transfer reactions, radical recombination, scavenging processes, ionic reactions in solution, to name but a few \cite{Rice1985, Ben-Avraham2005}.

For reaction-diffusion processes taking place in low dimensional supports, the mean-field assumption becomes questionable, since it relies on the suitability of local classical rate equations. The latter implicitly assume a good local mixing of the reactants and are expected to fail in low dimensions as a result of the poor mixing induced by the restricted geometry of the embedding support. In this case, microscopic fluctuations are gradually amplified and end up by inducing strong deviations from the mean-field prediction. In systems with short-range interactions and low or negligible particle mobility, the consequences of such deviations are even more drastic \cite{Privman1993}. In this context, diffusion-limited chemical reactions fall under the category of systems whose particle mobility is often insufficient to ensure the validity of the mean-field picture. In order to obtain a more correct picture, a wide array of alternative approaches accounting for the role of fluctuations have been developed.

In the above context, the kinetics of seemingly simple systems is often far more intricate than expected. Elementary systems have the advantage that the influence of each parameter can be more easily elucidated, even if the resulting kinetics is not always obvious. Here, we shall revisit two standard reaction-diffusion systems, namely, the (irreversible) single-species coalescence reaction, A+A$\to$ A  as well as a close relative of this process, i.e., the annihilation reaction, A+A$\to \emptyset$. In the diffusion-limited case, these two binary reactions are known to share the same universality class \cite{Peliti1986}, and they are relevant to understand many experimental systems, ranging from diffusion-limited aggregation to annihilation of photoexcited solitons \cite{Ben-Avraham2005}.

The above reactions have been comprehensively studied in static media, especially in the decades around the last turn of the century, and are still a subject of active research \cite{Winkler2012,Allam2013,Fortin2014,Durang2011,Durang2014,Matin2015}.
A plethora of different methods have been developed to investigate the behavior, notably, scaling arguments \cite{Ben-Avraham2005}, the method of the interparticle distribution function (IPDF) \cite{Ben-Avraham1995b,Ben-Avraham2005}, the even/odd interval method \cite{Masser2000,Masser2001,Masser2001a}, renormalization group techniques \cite{Peliti1986,Lee1994,Winkler2012}, hierarchies of $n$-point distribution functions \cite{Lin1990,Lindenberg1995}, as well as other alternative methods \cite{Torney1983,Spouge1988,Sancho2007,Durang2014}.
Scaling arguments based on physical intuition are amazingly powerful given their simplicity, and correctly predict the decay exponent in the asymptotic long-time regime. However, such approaches fail to capture the subtleties of the process, for which one must resort to alternative techniques such as the IPDF method \cite{Doering1989} or its equivalent formulation in terms of empty interval probabilities \cite{Ben-Avraham1995b,Ben-Avraham2005}. This method yields exact results for the one-dimensional case and has the advantage of remaining suitable when additional reactions come into play, e.g. particle birth and/or the back reaction (reversible case) \cite{Doering1989,Burschka1989}. Other situations which also lend themselves to treatment via the IPDF/interval method involve finite systems \cite{Mandache2000}, inhomogeneous systems \cite{Doering1991,Krapivsky2015}, the inclusion of traps \cite{Ben-Avraham1998b}, the study of correlation functions beyond the two-point case \cite{Ben-Avraham1998c}, and the inclusion of a resetting process \cite{Durang2014}. One may therefore say that the coalescence- and annihilation-diffusion processes in static domains are, in many instances, well characterized and understood from a theoretical point of view. Nevertheless, a word of caution is in order here, since the majority of studies primarily focus on the long-time behavior and devote less attention to the transient regime.

The aforementioned results for the A+A reactions hold for the standard case of static domains. However, given the relevance of expanding media (notably in cosmology and developmental biology), one may wonder how the kinetics of simple reaction-diffusion systems is affected by the expansion of the medium in which they take place. As it turns out, the mixing properties of diffusing particles are strongly affected by the medium expansion (contraction) if the latter is sufficiently fast \cite{Simpson2015,Simpson2015b,Yuste2016,LeVot2017}. In this regime, the impact of the initial condition on the subsequent dynamics is enhanced by the medium expansion, which also has an influence on the well-known depletion effects associated with the chemical reactions.  One therefore anticipates that the kinetics of diffusion-controlled reactions will be strongly affected by dilution effects associated with the medium expansion. This is indeed confirmed by many studies concerning a number of reaction-diffusion systems in growing domains. Such studies concern both theory and applications, with special emphasis on biological systems \cite{Kulesa1996,Crampin1999,Crampin2001,Murray2003,Madzvamuse2010,Baker2010,
Simpson2015b} and related phenomena such as morphogen gradient formation \cite{Averbukh2014,Fried2015}. On much larger scales, the assumption of an
\emph{expanding} universe is required for the study of key processes such as
diffusion of cosmic rays in the extragalactic space \cite{Berezinsky2006,Berezinsky2007,Kotera2008,Batista2014}, structure formation in the universe driven by aggregation processes  \cite{Silk1978}, and annihilation of antimatter before or during the big bang nucleosynthesis \cite{Rehm1998}.

Returning to the specific case of the diffusion-limited A+A$\to$A and A+A$\to \emptyset$ reactions in one dimension, one may ask how their well-known phenomenology changes if one allows for a medium expansion, and, specifically,  whether the IPDF method remains a suitable tool to study such changes. In a very recent work about the kinetics of the annihilation reaction on a $1d$ domain which grows according to a power law  \cite{OurPreprint2018}, we have (partially) answered the second question in the affirmative. In the present work we carry out a detailed analysis of the coalescence reaction by means of the IPDF method (conveniently generalized to deal with growing media), and also extend the study of the annihilation reaction performed in Ref.~\cite{OurPreprint2018} along these lines. More specifically, we give the exact solution for the coalescence reaction on a growing $1d$ domain. In doing so, we do not limit ourselves to the behavior of the global particle concentration (as done in Ref.~\cite{OurPreprint2018} for the special case of the $1d$ annihilation reaction with a power-law domain growth and a random initial condition), but also consider the influence of the type of domain growth and initial condition on the particle concentration and on the associated survival probability, as well as the behavior of correlation functions characterizing the onset of spatial ordering in the system (in all cases, exact expressions for the full spatial and temporal dependence are derived). While the obtained expressions are similar to those for the static case, in the case of a sufficiently fast domain growth, the physical behavior may become very different. In particular, a non-empty final state is attained, whereby the associated survival probability depends strongly on the details of the initial condition.

The remainder of this paper is organized as follows. In Sec.~\ref{IPDFcoa}, we show in detail how to extend the IPDF method for the coalescence reaction to the case of an arbitrarily growing domain. In Sec.~\ref{secCoagu}, we show how the equations arising from the generalized IPDF method for coalescence reactions are readily solved via a suitable transformation in time and space.
We use these results to comprehensively discuss the behavior of the survival probability and of the particle concentration both analytically and by means of numerical simulations. In this context, a scaling analysis reproducing the main features of the kinetics is also given. We then go on to study properties characterizing the spatial ordering induced by the reactions in all time regimes, thereby paying special attention to the so-called IPDF function. In Sec.~\ref{secAniqui}, we discuss the behavior of the annihilation reaction on a uniformly growing domain. As already mentioned, in this paper we go well beyond previous results that were restricted to the concentration in the special case where the time growth of the medium is described by a power law \cite{OurPreprint2018}. Here, we consider the case of an arbitrary time growth of the medium, and discuss the behavior of the concentration and of the IPDF (as well as of the closely related two-point correlation function) over the whole time domain. The obtained results are also confirmed by numerical simulations. Finally, in Sec.~\ref{secConc}, we give a summary of the main conclusions and outline possible avenues for future research.

\section{IPDF method for the coalescence reaction on a growing 1d domain}
\label{IPDFcoa}

The IPDF method for the coalescence reaction $A+A\to A$  was originally developed in Refs.~\cite{Doering1988,Doering1989,Burschka1989,Doering1990}.
A detailed description thereof can be found in Refs.~\cite{Ben-Avraham1995b,Ben-Avraham2005}.
Assume that we have an infinite line populated by a collection of randomly scattered, Brownian point particles, each of them with the same diffusivity $D$. The global initial particle concentration (average number of particles per unit length) is $c_0$. Let $Y(t)$ denote the distance between a given particle, say $A_n$, and the particle $A_{n+1}$ to its right. Let us further assume without real loss of generality that $A_n$ disappears instantaneously when it collides with $A_{n+1}$, but survives when it encounters $A_{n-1}$. In this case, the probability density function $\hat\sigma(y,t)$  associated with the interparticle distance $Y(t)$ obeys the diffusion equation of a Brownian particle in the reference frame of its nearest neighbor, that is,
\begin{equation}
\label{sigmaecu}
\frac{\partial   \hat\sigma}{\partial t}=2D\frac{\partial^2 \hat \sigma}{\partial y^2},\quad y>0
\end{equation}
(note that the diffusivity then becomes twice as large, i.e., equal to $2D$). The above equation must be complemented with an absorbing boundary condition at the origin $\hat\sigma(0,t)=0$ (see more details in Ref.~\cite{Doering1988}).  The survival probability $S(t)$ of a randomly chosen particle after a time $t$ is simply $S(t)=\int_0^\infty \hat\sigma(y,t) dy$.  Introducing the auxiliary function $\hat q(y,t)=  c_0\hat\sigma(y,t)$, the concentration of particles after a time $t$ is obtained as follows:
\begin{equation}
\label{comconc}
\hat c(t)=  c_0 S(t)=\int_0^\infty  \hat q(y,t) dy.
\end{equation}
The interparticle probability density function (IPDF), $\hat p(y,t)$, defined as the density of the probability of finding a gap of size $y$ between neighboring particles is the key function of the IPDF method. It is given by \cite{Doering1988}
\begin{equation}
\hat p(y,t)=\hat q(y,t)/\hat c(t).
\end{equation}
From the definition $\hat q(y,t)=c_0\hat\sigma(y,t)$ and from Eq.~\eqref{sigmaecu}, one sees that $\hat q(y,t)$ satisfies the diffusion equation
\begin{equation}
\label{hatqecu}
\frac{\partial  \hat q}{\partial t}=2D\frac{\partial^2 \hat q}{\partial y^2},\quad y>0
\end{equation}
and the boundary conditions $\hat q(0,t)=\hat q(\infty,t)=0$. The boundary condition $\hat q(0,t)=0$ reflects the fact that the distance between two neighboring particles can never become zero, since one of them will vanish upon encounter; the second boundary condition $\hat q(\infty,t)=0$ states that no infinite gap exists in the system as long as the number of particles is larger than one.

Another important quantity is the probability $\hat E(y,t)$ for a randomly chosen interval of length $y$ to be devoid of particles \cite{Ben-Avraham1995b,Ben-Avraham2005}.  This quantity can be obtained from the IPDF by integrating $\hat q(y,t)$ twice over the spatial variable:
\begin{equation}
\label{Eyt}
\hat E(y,t)=\int_0^y dy' \int_0^{y'} dy'' \hat q(y'',t)-\hat c(t) y +1,
\end{equation}
which implies
\begin{equation}
\label{cfromE}
\hat c(t)=-\left. \frac{\partial \hat E}{\partial y}\right|_{y=0}
\end{equation}
 and
\begin{equation}
\label{pfromE}
\hat q(y,t)=\hat c(t) \hat p(y,t)= \frac{\partial^2 \hat E}{\partial y^2}.
\end{equation}
The empty interval probability $\hat E(y,t)$ satisfies the same diffusion equation as $\hat \sigma(y,t)$, but subject to different boundary conditions: $\hat E(0,t)=1$ and $\hat E(\infty,t)=0$ \cite{Ben-Avraham1995b,Ben-Avraham2005}.   In fact, an alternative way to compute $\hat c(t)$ and $\hat p(y,t)$ is to directly obtain $\hat E(y,t)$ by solving this boundary value problem and then using Eqs.~\eqref{cfromE} and \eqref{pfromE}.

\begin{figure}[t]
\begin{center}
        \includegraphics[width=0.4\textwidth,angle=0]{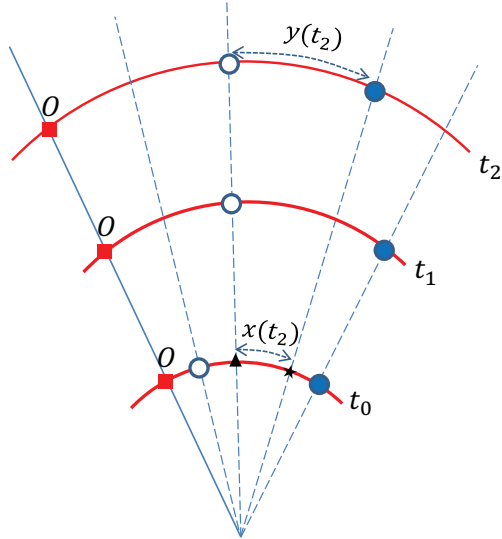}
\end{center}
\caption{\label{FigComovil}
Relationship between the comoving and physical coordinates of two diffusing particles (open and filled circles) at three different times, $t=0$, $t=t_1$, and  $t=t_2>t_1$. Both particles move on a one-dimensional growing domain. Solid arcs represent the physical domain at the corresponding times.  During the interval $(0,t_1)$ the particle to the left (open circle) performs a diffusive jump to a  position whose comoving coordinate is marked by a triangle, whereas the right particle (solid circle) does not jump.   During the time interval $(t_1,t_2)$ the first particle does not jump and therefore its comoving coordinate does not change; during the same interval, the second particle jumps to a position whose comoving coordinate is marked by a star. The comoving distance $x(t_2)$ corresponding to the physical distance $y(t_2)$  between the two particles at time $t_2$ (i.e., arc distance labeled as $t_2$) is just the distance between the star and the triangle.}
\end{figure}

Let us now consider how the dynamics of the stochastic variable $Y(t)$ changes if one allows for a (deterministic) expansion of the spatial domain (in our case, a line).  The (physical) coordinate $y$ of the Brownian particle with diffusivity $2D$ (i.e., the distance $y$ between two neighboring particles as defined at the beginning of this section) will change even if it has no intrinsic motion (i.e., even if neither of the two neighboring particles performs Brownian jumps). The reason is that the particle still experiences a deterministic drift induced by the expansion of the domain. The associated drift velocity will be hereafter denoted by $u(y,t)$. In the language of cosmology, the particle is drifted with velocity $u(y,t)$ by the so-called Hubble flow \cite{Perkins2008}. A standard method to simplify the theoretical description of this expansion process consists in switching to comoving coordinates. These coordinates are defined as follows (see also Fig.\ref{FigComovil}). The one-dimensional domain on which the particles move can be seen as a continuum of ``fixed points'', i.e., points lacking intrinsic motion. Obviously, at any time $t$,  the diffusive particle occupies the position $y$ of a fixed point. The position $x$ of this fixed point at $t=0$ defines the comoving coordinate of the particle at time $t>0$. From this definition one sees that $y(0)=x$.  In general, the relationship between the physical coordinate $y$ and the corresponding comoving coordinate $x$ can be written as $y(t)=f(x,t)$, where $f$ is a function that is continuous in both arguments and bijective in $x$ for any $t\ge 0$. A special yet important case is given by the equation $y(t)=a(t) x$, which corresponds to the case of a uniform expansion. Note that, by definition, $a(0)=1$.  In cosmology, the coefficient $a(t)$ is known as the scale factor. In this case, the associated drift velocity is $u= \dot{a} x=(\dot{a}/a) y$.

On a growing domain, the diffusion equation of a Brownian walker is identical with a standard diffusion equation augmented with an additional drift-dilution term describing the effect of the domain growth \cite{Crampin1999,Yuste2016}. Thus, Eq.~\eqref{sigmaecu} for $\hat \sigma(y,t)$, or equivalently Eq.~\eqref{hatqecu} for $\hat q(y,t)$, becomes
 \begin{equation}
\label{rhoExpan}
\frac{\partial \hat q}{\partial t}=2D\frac{\partial^2 \hat q}{\partial y^2} - \frac{\partial (u \hat q)}{\partial y}
\end{equation}
as a result of the expansion process. In the above equation, the additional term introduced by the medium growth (last term on the right hand side) can be split into two contributions with a clear physical interpretation, namely, a drift term $-u\partial \hat q/\partial y$, and a dilution term $-\hat q \partial u/ \partial y$ \cite{Crampin1999,Yuste2016}. The domain growth does not modify the boundary condition for $\hat q$, i.e., one still has $\hat q(0,t)=\hat q(\infty,t)=0$.

\section{Exact solution for the coalescence reaction on a 1d growing domain}
\label{secCoagu}

\subsection{General formalism in terms of comoving coordinates}
\label{secCoaguGF}
In general, solving Eq.~\eqref{rhoExpan} for arbitrary growth processes is a challenging task. However, it is possible to obtain a general solution for the case of a homogeneous expansion, $y(t)=a(t) x$.  To this end, it is convenient to switch to comoving coordinates. Let $X(t)$ be the stochastic variable describing the comoving coordinate of the particle corresponding to its physical coordinate $Y(t)$.   The relationship between the probability density functions $\sigma(x,t)$ and  $\hat \sigma(y,t)$ respectively associated with $X(t)$ and $Y(t)$ is $\hat \sigma(y,t)dy=\sigma(x,t)dx$, which simply states that probability is conserved.  For uniform domain growth one has $\hat\sigma(y,t) =\sigma(x,t)/a(t)$ or, equivalently, $\hat q(y,t)=q(x=y/a(t),t)/a(t)$ where $q(x,t)=c_0 \sigma(x,t)$.   Plugging this latter expression into Eq.~\eqref{rhoExpan}, one obtains \cite{Yuste2016,LeVot2017}
 \begin{equation}
\label{rhoExpanComov}
\frac{\partial q}{\partial t}=\frac{2D}{a(t)^2}\,\frac{\partial^2 q}{\partial x^2}.
\end{equation}
The boundary conditions $\hat q(0,t)=\hat q(\infty,t)=0$ imply $q(0,t)=q(\infty,t)=0$.
Formally, Eq.~\eqref{rhoExpanComov} is a diffusion equation with a time-decreasing diffusivity. The reason for this is clear: In comoving space, Brownian jumps become increasingly shorter as a result of the fact that the jump size distribution in physical space is not coupled to the expansion of the domain. Consequently, a point at a given initial distance from the walker's starting position becomes increasingly harder to reach.
More precisely, after a given time $\delta t$, the mean square displacement of a particle in comoving space is $(\delta x)^2=2 (D/a(t)^2) \delta t$.
Since the diffusivity is proportional to the variance of the single-jump displacement, we conclude that the typical length of the latter is shortened by a factor $1/a(t)$ with respect to the same displacement measured in physical space. In passing, we note that this description in terms of shortened displacements in comoving space is a very natural way to approach the problem of anomalous diffusion in growing domains \cite{LeVot2017,Angstmann2017}. Note that the time evolution of the physical interparticle distance $Y(t)$ can be formulated in terms of its comoving counterpart $X(t)$ as a diffusion process taking place on a \emph{static} domain, at the expense of having to deal with a time-dependent diffusivity.

Using the same argument as in Sec.~\ref{IPDFcoa}, one has
\begin{equation}
\label{ctSt}
c(t)=  c_0 S(t)=\int_0^\infty  q(x,t) dx,
\end{equation}
where $c(t)$ is the number density of particles in comoving space (number of particles per unit length of the static domain).
Similarly, the IPDF in comoving coordinates, $p(x,t)$, is given by
\begin{equation}
\label{pxt}
 p(x,t)= q(x,t)/ c(t).
\end{equation}

The probability $E(x,t)$ that a randomly chosen interval of comoving length $x$ is devoid of particles can also be obtained from the IPDF by integrating $  q(x,t)$ twice over the spatial variable [cf. Eqs.~\eqref{Eyt}-\eqref{pfromE}]:
\begin{equation}
\label{Ext}
 E(x,t)=\int_0^x dx' \int_0^{x'} dx''q(x'',t)- c(t) x +1,
\end{equation}
which implies $  c=-\partial   E/\partial x|_{x=0}$ and $  q(x,t)=   c(t)  p(x,t)= \partial^2   E/\partial x^2$.

\subsection{Brownian conformal time}
In order to obtain $q(x,t)$ explicitly,  we shall first cast Eq.~\eqref{rhoExpanComov} into a diffusion equation with constant diffusivity. To this end, let us
introduce the new variable $\tau(t)$ \cite{Yuste2016},  defined via the equation
\begin{equation}
\label{ConfTimeDef}
\tau(t)=\int_{0}^t \frac{ds}{a^2(s)}.
\end{equation}
We term $\tau(t)$ Brownian conformal time by analogy with the standard conformal time $\tau_c(t)=\int_{0}^t ds/a(s)$ used in cosmology. In terms of $\tau(t)$, Eq.~\eqref{rhoExpanComov} can be expressed as
 \begin{equation}
\label{rhoExpanComovTau}
\frac{\partial q}{\partial \tau}=2D\,\frac{\partial^2 q}{\partial x^2},
\end{equation}
i.e., a standard diffusion equation with constant diffusivity.

For convenience, we shall occasionally switch between different notations for the same functions from now on, e.g., $q(x,t)$ and $q(x,\tau)$, or $S(t)$ and $S(\tau)$, etc. The connection between both notations is, of course, $q(x,t)=q(x,\tau(t))$, $S(t)=S(\tau(t))$, etc.

In passing, we note that the definition of the Brownian conformal time opens the door to a non-standard interpretation of the mean square displacement of the random walk. This quantity can indeed be written as $(\delta x)^2=2 D (\delta t/a(t)^2)= 2D \delta \tau (t)$, i.e., the diffusivity takes the same (constant) value as in physical space, but the time flow in the static (comoving) domain is slowed down. This is easily seen by differentiating Eq. \eqref{ConfTimeDef} with respect to the physical time:
\be
\label{difftau}
\frac{d\tau}{dt}=\frac{1}{a(t)^2}.
\ee
The behavior of the conformal time $\tau(t)$ is  key for understanding the reaction kinetics on the original growing domain. In terms of the domain growth, one can distinguish three different regimes for which it is useful to coin a proper terminology.

We shall hereafter refer to the domain growth or expansion as ``over-Brownian'' when $a(t)$ increases in time at a large enough rate, so as to ensure that $\tau(t)$ tends to a finite value at long times, i.e., $\tau(\infty)\equiv \tau_\infty<\infty$ as $t\to\infty$. In particular, this holds when $a(t)\sim t^\gamma$ with $\gamma>1/2$. In this case the integral in Eq.~\eqref{ConfTimeDef} converges and, consequently, $\tau(t)$ goes to a finite value, $\tau_\infty$. The use of the term ``over-Brownian expansion'' in the above case of a power-law expansion with $\gamma>1/2$ is justified by the fact that the distance between two fixed points [which is proportional to $a(t)$]  grows faster than the root-mean-square displacement of a Brownian random walker [which is known to increase as $t^{1/2}$].  In the opposite case of a sufficiently slow expansion ($\gamma<1/2$), the latter will be said to be under-Brownian; in this case $\tau(t) \to \infty$ for $t\to \infty$. For $\gamma=1/2$, we will speak of a Brownian expansion (not in the sense that the expansion is stochastic, but rather than in the sense that the distance between two fixed points grows as fast as the root-mean-square-displacement of a Brownian particle). Still, we have a fourth possibility, that is, $a(t)$ grows faster than $t^{1/2}$ but still $\lim_{t \to \infty} \tau(t)=\infty$; we will refer to this case as ``marginally
over-Brownian''. Table \ref{tablaExptypes} gives an overview of the four different types of expansion.
\begin{table}
\begin{center}
    \begin{tabular}{ | l | l | c |}
    \hline
    Expansion & Asymptotic behavior & Asymptotic Brownian conformal time \\ \hline
    Under-Brownian & $\lim_{t \to \infty} a(t)/\sqrt{t}=0$ & $\lim_{t \to \infty} \tau(t)= \infty$ \\ \hline
    Brownian & $\lim_{t \to \infty} a(t)/\sqrt{t} \in (0,\infty) $& $\lim_{t \to \infty} \tau(t)= \infty $\\ \hline
    Marginally over-Brownian & $\lim_{t \to \infty} a(t)/\sqrt{t}= \infty $& $\lim_{t \to \infty} \tau(t)= \infty $\\ \hline
    Over-Brownian & $\lim_{t \to \infty} a(t)/\sqrt{t}=\infty$ & $\lim_{t \to \infty} \tau(t) < \infty$ \\
    \hline
    \end{tabular}
\end{center}
\caption{\label{tablaExptypes} Classification of the four different types of expansion.}
\end{table}
For the sake of simplicity, we will not consider expansions that do not fall within
any of the above cases, such as expansions for which the asymptotic behavior
is not well defined.

As we shall see, over-Brownian expansions lead to drastic changes in the kinetics of the considered diffusion-reaction processes; in particular, they lead to an eventual freeze-out scenario where the ongoing encounter-controlled reactions prematurely come to a halt.

Two important cases of uniform expansion are the case of a power-law,
$a(t)= (1+t/t_0)^\gamma$, and the exponential case,  $a(t)=\exp(Ht)$. Both types of expansion are often found in the context of cosmology: $a(t)\sim t^{1/2}$  describes a  radiation-dominated universe, $a(t)\sim t^{2/3}$ describes a  matter-dominated universe, and $a(t)\sim \exp(Ht)$ corresponds to a dark-energy-dominated universe, $H$ being the Hubble parameter \cite{Ryden2003}. In passing, we also note that systems involving exponentially growing media are very common in developmental biology \cite{Kulesa1996,Binder2008,Murray2003}. For a power-law expansion, the Brownian conformal time takes the form
\begin{subnumcases}{\label{powerlawconftime} \tau(t)=}
\label{powerlawconftimea} \dfrac{t_0}{2 \gamma - 1} \left[ 1 - \left( \dfrac{t+t_0}{t_0} \right)^{1 - 2\gamma }  \right]  &   if \;\; $2\gamma \neq 1$, \\[2mm]
 t_0 \ln \left( \dfrac{t+t_0}{t_0} \right)  &   if \;\; $2\gamma = 1 $,
\end{subnumcases}
 whereas in the case of an exponential expansion one obtains
\begin{equation}
\label{tauExp}
\tau(t)=\frac{1- \exp(-2H t)}{2 H}.
\end{equation}
We thus have that $\tau(t)$ goes to the finite values $\tau_\infty=t_0/(2\gamma-1)$
when  $\gamma>1/2$ and $\tau_\infty=1/2H$ for $H>0$. Another kind of expansion (of special relevance in biology) is the case of a logistic growth \cite{Crampin1999,Baker2010}:
\begin{equation}
a(t)=\frac{a_\infty e^{\alpha t}}{ a_\infty-1+e^{\alpha t} },
\end{equation}
with $\alpha, a_\infty >0$ and where $a_\infty < 1$ describes a (logistic) contraction, $a_\infty > 1$ describes a (logistic) medium expansion, and $a_\infty=1$ simply describes a static medium. In this case
\begin{equation}
\label{tauLogi}
\tau(t)=\frac{1}{2\alpha a_\infty^2}
\left[ 2\alpha t+(a_\infty+1)^2-4-4 (a_\infty-1)e^{-\alpha t} -(a_\infty-1)^2 e^{-2\alpha t} \right].
\end{equation}
For long times one finds $\tau(t)\sim t/a_\infty^2$,  i.e.,  $\tau$ ends up growing linearly in time as in the static case. Thus, one expects that, up to subdominant transient effects, the long-time behavior of the system will essentially be given by that of the static case, independently of the parameter choice.

Coming back to the boundary value problem defined by Eq.~\eqref{rhoExpanComovTau} and the boundary conditions $q(0,\tau)=q(\infty,\tau)=0$, one immediately sees that the equations are identical with those for the static case [cf. Eq.~\eqref{hatqecu}], except for the fact that $t$ and $y$ are now replaced with the transformed variables $\tau$ and $x$. This means that the kinetics of any relevant quantity can be readily obtained by performing the double replacement $t\to \tau$ and $y\to x$ in the expression for its counterpart in the static case. In what follows, we shall drop the hat symbol when referring to quantities for the non-static case expressed in the ($x$, $\tau$) representation, e.g.,  $\hat q(y,t)\to q(x,\tau)$ or $\hat c(t) \to c(\tau)$.

A central quantity in the IPDF approach is the Green's function associated with $q(x,t)$ \cite{Ben-Avraham1990},
\begin{equation}
\label{Gxx}
G(x,x',\tau)=\frac{1}{(8\pi D \tau)^{1/2}} \,
\left\{
\exp\left[-\frac{(x-x')^2}{8D\tau}\right]
-\exp\left[-\frac{(x+x')^2}{8D\tau}\right]
\right\},
\end{equation}
i.e., the solution of the boundary value problem defined via Eq.~\eqref{rhoExpanComovTau}, the boundary conditions $q(0,\tau)=q(\infty,\tau)=0$, and the initial condition $q(x,0)=\delta(x-x')-\delta(x+x')$. For $x>0$, the Green's function can be interpreted as the probability density function of a Brownian particle on the positive half-line subject to the action of an absorbing boundary at $x=0$, and given that this particle starts at $x'>0$ \cite{Redner2001}.
Knowledge of the Green's function allows one to obtain the generic solution $q(x,\tau)$ for any initial condition $q(x,0)=p(x,0)/c_0$. One has
\begin{equation}
\label{qInt}
q(x,\tau)=\int_0^\infty dx' G(x,x',\tau) q(x',0).
\end{equation}
The corresponding survival probability $S(t)$ of the reacting particles is then obtained from Eq.\eqref{ctSt} and reads
\begin{equation}
\label{StStau}
S(\tau)=\frac{1}{c_0}\int_0^\infty  q(x,\tau) dx.
\end{equation}
As in well-known case of a static domain, the above result can be regarded as a natural consequence of reducing the original multiparticle problem to a two-particle problem (or to a first-passage problem in the reference frame of one of the Brownian particles). This is precisely the property exploited by the IPDF method.

\subsection{Short- and large-$\tau$ asymptotics}

 In the standard case of a static domain, it is well known that memory effects are strongly weakened in the long-time regime, as a result of which the concentration $c(t)$ becomes independent of the initial condition (except for certain ``exotic cases'' such as fractal or power-law initial distributions \cite{Alemany1997a,Mandache2000}). However, the short-time behavior is largely influenced by the initial arrangement of the particles.

 In the case of an expanding domain, an analogous discussion can be carried out not in terms of the physical time $t$, but rather in terms of the conformal time $\tau$. This distinction is in some cases important because $\tau(t)$ is, in general, a non-linear function of $t$. Thus, in those cases where $\tau(t)$ always remains small as $t$ grows, the small-$\tau$ expressions obtained by performing the replacement $t\to \tau$ in the small-$t$ expressions for the static case remain valid even for very long times. This e.g. holds for power-law and exponential expansions when $\gamma$ and $H$ are very large, respectively.  Conversely, when $\tau(t)$ grows very fast, the large $\tau$ expressions turn out to be valid already for short times $t$.  In particular, this is the case for fast contracting media.

 For the case of an uncorrelated random initial distribution of particles (i.e., a Poisson distribution),  the initial IPDF is $p(x,0)=c_0\exp(-c_0 x)$. In the small $\tau$ regime one finds \cite{Doering1988}
\begin{equation}
\label{RanICSt}
S(\tau) = 1- (8c_0^2 D \tau/\pi)^{1/2}+O(c_0^2 D \tau).
\end{equation}
In contrast, for a periodic configuration $p(x,0)=\delta(x-c_0^{-1})$, one has
\begin{equation}
\label{PerICSt}
S^\text{per}(\tau) = 1- (8c_0^2 D \tau/\pi)^{1/2} \exp(-1/8c_0^2 D \tau)\left[1+O(c_0^2 D\tau)\right].
\end{equation}
The form of the subleading term in the above cases gives a rough idea of the time regime in which the leading term is expected to provide a reasonable approximation, namely, a time interval for which the condition $c_0^2 D\tau(t)\ll 1$ holds. In addition, in all cases we tacitly assume that the value of the initial concentration $c_0$ is low enough, so that most of the particles experience a truly diffusive process before reacting. This avoids transient deviations from the prediction of the diffusion equation in the short-time regime. For a Poissonian initial distribution, the survival probability of the particles at a given (sufficiently short) time is smaller than for the periodic distribution  because in the former case there is a non-negligible probability of finding particles very close to each other.

On the other hand, if the expansion is such that $\tau(t)$ reaches large values as $t\to\infty$ (which is always possible when under-Brownian, Brownian, and
marginally over-Brownian expansions are considered), the concentration and the IPDF describing the spatial ordering of the particles become independent of the initial condition. Using the long-time form
\begin{equation}
G(x,x',\tau)\approx  \frac{2x'}{(8\pi D \tau)^{1/2}} \, \frac{2x}{8D\tau}
\exp\left(-\frac{x^2}{8D\tau}\right)
\end{equation}
of the Green's function \cite{Ben-Avraham1990} in Eqs. \eqref{pxt}, \eqref{qInt}, and \eqref{StStau}, one finds
\begin{equation}
\label{ctAsin}
c(\tau)\approx \frac{1}{ (2\pi D\tau)^{1/2}}
\end{equation}
and
\begin{equation}
\label{pxtAsin}
p(x,\tau)\approx  \frac{x}{4 D \tau} \, \exp\left(-\frac{x^2}{8D\tau}\right)
\end{equation}
for large $\tau$.

\subsection{Solution for a completely random initial distribution}

For a Poissonian initial distribution one has $q(x,0)=c_0p(x,0)=c_0^2 \exp(-c_0 x)$, implying that the integral on the right hand side of Eq.~\eqref{qInt} [with $G(x,x',\tau)$ given by  Eq.~\eqref{Gxx}] can be explicitly computed:
\begin{equation}
q(x,\tau)=\frac{c_0^2}{2}\,  e^{2c_0^2 D \tau}\,
\left\{
e^{-c_0 x} \text{erfc}\left[-\frac{x-4c_0 D\tau}{(8D\tau)^{1/2}}\right]
-
e^{c_0 x} \text{erfc}\left[\frac{x+4c_0 D\tau}{(8D\tau)^{1/2}}\right]
\right\}
\end{equation}
whence, by virtue of \eqref{ctSt}, we obtain
\begin{equation}
\label{Stazar}
S(t)= e^{z^2}\text{erfc}(z),
\end{equation}
with $z=c_0\sqrt{2 D \tau(t)}$.  From the above two equations one obtains the exact IPDF $p(x,t)=q(x,t)/[c_0S(t)]$ for a initial Poisson distribution of particles:
\begin{equation}
\label{pxtPoisson}
p(x,\tau)=\frac{c_0}{2 }\,
\left\{
e^{-c_0 x} \text{erfc}\left[-\frac{x-4c_0 D\tau}{(8D\tau)^{1/2}}\right]
-
e^{c_0 x} \text{erfc}\left[\frac{x+4c_0 D\tau}{(8D\tau)^{1/2}}\right]
\right\}\Big/\text{erfc}\left( \sqrt{2c_0^2 D \tau}\right).
\end{equation}

It is also possible to obtain the analytical expression of $E(x,\tau)$ by carrying out the corresponding integrals in Eq.~\eqref{Ext}:
\begin{align}
\label{Extau}
E(x,\tau)&=
\text{erfc}\left(\frac{x}{\sqrt{8D\tau}}\right)-\frac{1}{2}e^{2Dc_0^2\tau} \nn \\
& \times \left\{ e^{c_0 x} \left[1-\text{erf}\left(\frac{x+4Dc_0\tau}{\sqrt{8D\tau}}\right) \right]-e^{-c_0 x} \left[1+\text{erf}\left(\frac{x-4Dc_0\tau}{\sqrt{8D\tau}}\right) \right]\right\}.
\end{align}
As already anticipated, the above theoretical expression follows directly from that for the static case \cite{Doering1988,Ben-Avraham1990,Ben-Avraham1998c,Masser2001a,Ben-Avraham2005}
by performing the simple replacements  $y\to x$ and $t\to \tau$.

On a growing domain, $\tau(t)$ displays sublinear behavior in $t$  and the survival probability is larger than in the static case. Correspondingly, the time decay of the concentration becomes slower. Moreover, if the expansion is over-Brownian, $\tau(t)$ will saturate to a \emph{finite} value $\tau_\infty$  as $t\to\infty$, implying a drastic change in the qualitative behavior; in the long-time regime, a number of particles are not able to cover the increasing separation distance induced by the drift associated with the domain growth. As a result of this, they remain unreacted, and the survival probability attains a non-zero asymptotic value $S_\infty\equiv S(t\to\infty)$. This behavior is indeed not unexpected. A similar behavior is observed when reactions are localized on a domain boundary that is shifted by the growth process \cite{Simpson2015,Simpson2015b,Yuste2016}.
Equation~\eqref{Stazar} allows us to compute the exact value of $S_\infty$ for a Poissonian initial condition:
\begin{equation}
\label{Scota}
S_\infty= \exp(c_0^2 2D \tau_\infty) \, \text{erfc}\left(c_0\sqrt{2 D \tau_\infty}\right).
\end{equation}
In the over-Brownian case discussed above,  the reactant distribution in the $t\to \infty$  will tend to a frozen distribution characterized by the
IPDF~\eqref{pxtPoisson} with $\tau=\tau_\infty$. In this limit, the spatial arrangement of the particles significatively differs from the one given by Eq.~\eqref{pxtAsin} provided that $\tau_\infty$ is not too large.  In fact, for any initial distribution of particles, if the domain growth is so fast that $\tau_\infty$ becomes very small, then the distribution of particles will asymptotically tend to a distribution, in comoving space, that is very close to the initial one, and $S_\infty$ will be given by Eqs.~\eqref{RanICSt} and \eqref{PerICSt}  with $\tau=\tau_\infty$.  In particular, if $\tau_\infty$ is small enough, the number of reactions is extremely low, implying that $S^\text{per}(\tau_\infty)\approx 1$.

\subsubsection{Survival probability. Discussion and comparison with simulation results}

\begin{figure}[t]
\begin{center}
        \includegraphics[width=\ancho\textwidth,angle=0]{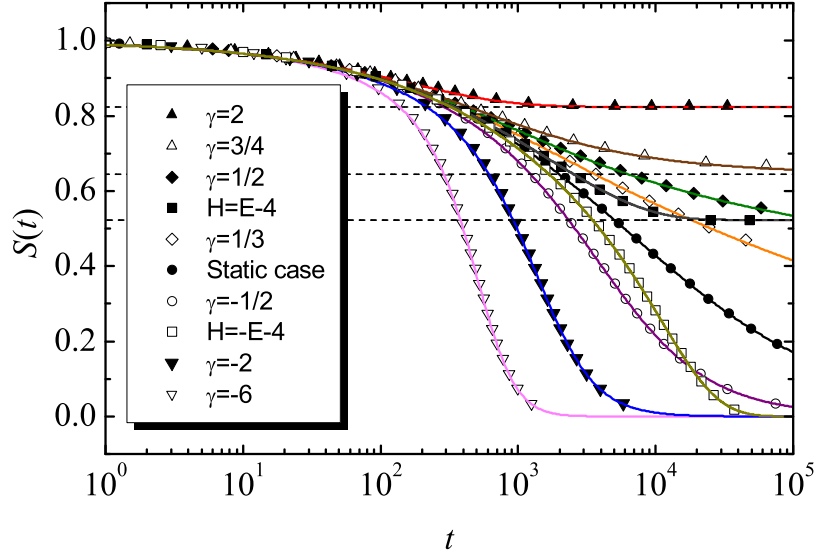}
\end{center}
\caption{\label{FigSttauLimCoagu}
Survival probability vs.~time for the coalescence reaction. Lines represent theoretical results [given by Eq.~\eqref{Stazar}] whereas symbols are simulation results. Here and in subsequent figures we use $10^5$ particles with $c_0=1/100$ and $D=1/2$. Eleven different cases are considered, i.e., power-law expansions with   $t_0=1000$ and $\gamma=\left\{2, 3/4, 1/2, 1/3, -1/2, -2, -6\right\}$, exponential expansions with $H=\left\{10^{-4},-10^{-4}\right\}$ , the logistic case with $a_\infty=3$ and $\alpha=10^{-4}$, and the static case $a(t)=1$. Dashed lines correspond to the asymptotic value $S_\infty$ for $\gamma=2$ ($S_\infty\approx 0.823$), $\gamma=3/4$ ($S_\infty\approx 0.644$) and $H=10^{-4}$ ($S_\infty\approx 0.523$).
}
\end{figure}

Displayed in Fig.~\ref{FigSttauLimCoagu} is a comparison between the theoretical expression of the survival probability $S(t)$ given by Eq.~\eqref{Stazar} for a completely random initial condition and the survival probability obtained from simulations for the cases of a power-law expansion and of an exponential expansion (including the case $a(t)<1$ of a contracting domain).  The agreement between the analytical results and the numerical results is excellent. For over-Brownian cases (cases with  $\gamma>1/2$ and $H>0$) the horizontal dashed line represents $S_\infty$, the asymptotic value of $S(t)$ given by Eq.~\eqref{Scota}.

The rich phenomenology illustrated by Fig.~\ref{FigSttauLimCoagu} can be explained as follows. For an over-Brownian expansion, gaps between neighboring particles grow quickly in physical space, and diffusive transport is not fast enough to sustain the encounter-controlled reactions. Specifically, since the span of the random walk grows as $t^{1/2}$, the above scenario will take place whenever $a(t)$ grows faster than $t^{1/2+\epsilon}$ with $\epsilon>0$ arbitrarily small. In particular, this is the case for any exponential expansion [$a(t) \propto e^{Ht}$ with $H>0$] and for any power-law expansion with $a(t) \propto t^\gamma$ and $\gamma>1/2$. As a result of this, in these cases (as well as in any other over-Brownian case) the survival probability tends to a finite value $S_\infty$, represented in Fig.~\ref{FigSttauLimCoagu} by dashed horizontal lines for different expansion processes.

In contrast, when $a(t)$ increases more slowly than $t^{1/2}$  (under-Brownian expansion), when $a(t)=1$ (static case) or when $a(t)$ decreases in time (case of a contracting domain),  the widening of interparticle gaps due to the domain growth is not enough to prevent particle encounters driven by diffusion. Consequently, any particle will sooner or later encounter another one, and reactions will take place until all the particles (but one) disappear, thereby leading to a zero macroscopic concentration and to a vanishing value of $S_\infty$. Conversely, when the expansion rate is decreased (or the contraction rate increased), the system becomes empty at a faster rate.

If one considers the reaction kinetics in comoving space, the above behavior can be explained as follows. Even though comoving distances remain constant, particle diffusivities decrease with $a(t)^{-2}$ [see Eq.~\eqref{rhoExpanComov} and the subsequent discussion], thereby leading to a decreased encounter rate. For a sufficiently fast expansion, fewer reactions take place and the survival probability tends to larger values. In the opposite limit of a sufficiently slow expansion [characterized by an unbounded growth of $\tau(t)$ as  $t\to\infty$] or in the case of a contracting domain, the survival probability goes to zero in the long-time limit.

Summarizing, the behavior of the conformal time $\tau$ will determine whether the reactions stop before the system becomes empty. If the expansion is over-Brownian,  $\tau$ will tend to a finite value $\tau_\infty$, and the typical comoving distance  traveled by the particles will also converge to a finite value $\ell=\sqrt{2D \tau_\infty}$ (the so-called Brownian event horizon, see Ref.~\cite{Yuste2016}). This means that, at low enough particle concentrations,  the typical interparticle distance, $1/c(t)$, will be larger than the typical distance $\ell$ a particle can ever reach. Therefore, a given particle will not collide with its nearest neighbor; consequently, the reaction will not take place, implying that $S_\infty>0$. In the limit where the concentration approaches $1/\ell$, one could once again borrow the language of cosmology and say that the reactions stop because the different particles evolve in separate ``Brownian universes'': the idea underlying this definition is that a given particle cannot reach any region outside its universe by means of a purely diffusive process. This statement holds of course only in a statistical sense, given that the possibility of arbitrarily long jumps induced by the corresponding Wiener process cannot be discarded. In contrast, when the expansion is under-Brownian, $\ell$ tends to infinity, encounters are certain, and consequently all the particles but one die
with certainty, implying that the macroscopic survival probability $S(t)$ will tend to 0 as $t\to \infty$.

Let us discuss the transient behavior in more detail. As the expansion rate decreases, the characteristic relaxation time needed to ``almost'' reach the asymptotic value first increases and then decreases again. For large expansion rates, very few reactions take place, and a nonzero value of $S_\infty$ quickly settles. For moderate expansion rates, a large number of reactions take place at intermediate times, as a result of which the characteristic time to reach the non-zero steady state increases. In contrast, for sufficiently low expansion rates, the majority of reactions take place at early times, and the time needed to reach the steady state (which, in this case, is an empty state) decreases with respect to the case of a moderate expansion rate.

The richness of the transient behavior is also illustrated by the fact that, in Fig.~\ref{FigSttauLimCoagu}, curves for different types of expansion may intersect each other. At short or intermediate times, a sufficiently slow exponential expansion may favor the reaction rate with respect to a power-law expansion, but at larger times the behavior is exactly the opposite.

\begin{figure}[t]
\begin{center}
        \includegraphics[width=\ancho\textwidth,angle=0]{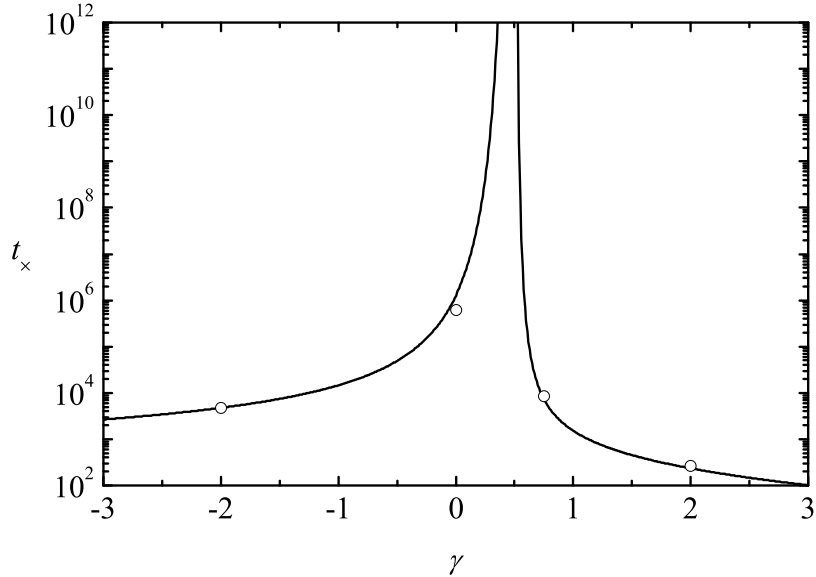}
\end{center}
\caption{\label{figtcross}
$t_\times$ vs.~ the exponent $\gamma$ for power-law expansions.  The line represents the theoretical value obtained by solving the equation $S(t_\times)=S_\infty+0.05$ with $S(t)$ and $S_\infty$ respectively given by Eqs.~\eqref{Stazar} and \eqref{Scota}.  The symbols represent simulation results. The chosen parameters are $t_0=1000$, $c_0=1/100$ and $D=1/2$.
}
\end{figure}

At this stage, a general remark is in order. The number of reactions per unit time is proportional to the negative time derivative $-dS/dt$ of the survival probability. In the static case, this quantity displays a maximum at a given characteristic time. For a given Poissonian initial condition, this non-monotonic behavior is also preserved on a growing domain, but the value of the characteristic time is shifted with respect to the static case. The behavior for different cases can be clearly seen in Fig.~\ref{figtcross}. Displayed in this figure is the $\gamma$-dependence of the relaxation time $t_\times$ defined by the equation $S(t_\times)=S_\infty+0.05$. In other words, $t_\times$ is the time at which the particle concentration differs from its asymptotic value $S_\infty$ by just five per cent. This time becomes very large (but is of course finite) in the vicinity of $\gamma=1/2$. For visualization purposes, the corresponding peak has been truncated in the figure.

\begin{figure}[t]
\begin{center}
        \includegraphics[width=\ancho\textwidth,angle=0]{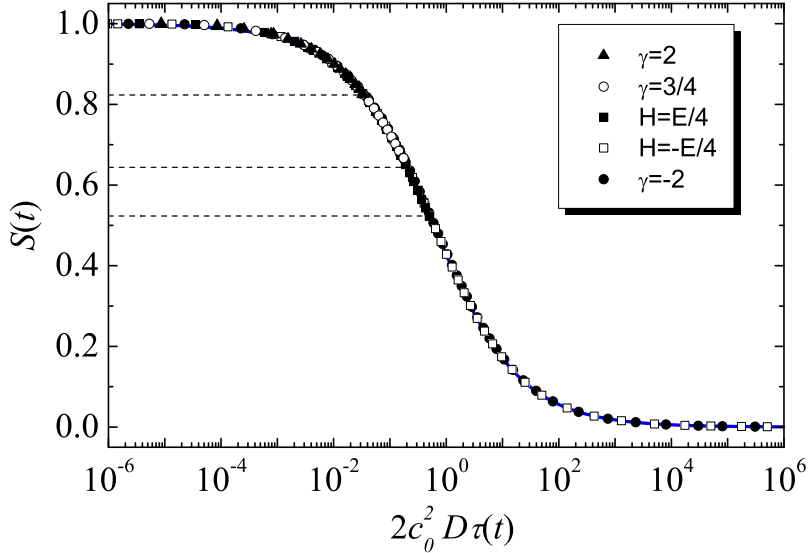}
\end{center}
\caption{\label{FigStCoagu}
Survival probability  $S(t)$ vs.~$z^2=2c_0^2 D \tau(t)$ for coalescence. The different symbols correspond to simulation results for power-law expansions with   $t_0=1000$ and $\gamma=\left\{2, 3/4,-2\right\}$, exponential expansions with $H=\left\{10^{-4},-10^{-4}\right\}$, and a logistic growth with $a_\infty=3$ and $\alpha=10^{-4}$.
Dashed lines correspond to the limiting value $S_\infty$ of the survival probability for $\gamma=2$ ($S_\infty\approx 0.823$), $\gamma=3/4$ ($S_\infty\approx 0.644$) and $H=10^{-4}$ ($S_\infty\approx 0.523$). The solid line corresponds to the exact solution, Eq.~\eqref{Stazar}.
}
\end{figure}

An alternative description of the reactive dynamics may be given in terms of the conformal time $\tau(t)$. Fig.~\ref{FigStCoagu} shows that curves with different expansion parameters fall onto a single universal curve when plotted in terms of the scaling parameter  $z^2=2c_0^2 D \tau(t)$. A physical interpretation of this scaling behavior is as follows. For a growing domain one has $a(t)>1$, implying that a small change in the conformal time $\delta \tau$ brings about a larger change $\delta t= a(t)^2 \delta \tau$ in physical time [cf. Eq.~\eqref{difftau}]. Thus, the slowing down of the diffusion-controlled reaction kinetics due to the domain growth is exactly counterbalanced by a slower flow of the conformal time, implying that the description of the time evolution of $S$ in terms of the new temporal variable $\tau$ is universal.

In spite of the above scaling behavior, in the over-Brownian case an important feature of $S(\tau)$ \emph{does} depend on the expansion rate, since here the survival probability goes to a finite value $S_\infty$. As already mentioned, this stems from the fact that the interparticle distance grows faster than the typical diffusive length. Effectively, this means that the concentration freezes at a non-zero value which depends on the expansion rate. Values of $S_\infty$ corresponding to different expansion rates are represented by the dashed horizontal lines in Fig.~\ref{FigStCoagu}.

While the values of $S_\infty$ displayed in Fig.~\ref{FigStCoagu} are all for the same initial condition, it is clear that, for a given type of expansion, the final value will depend on the initial condition. This behavior is at odds with the result for the static case, and reflects the strong memory effects of the initial distribution of particles characteristic of the over-Brownian case.

\subsection{Long-time concentration in physical space}

The reaction process A+A$\to$ A,  always tends to decrease the particle concentration $\hat c(t)$ in physical space; simultaneously, the growth process of the domain also tends to dilute the system due to the increased volume (the contrary is true in the case of a shrinking medium). This behavior is described by the equation
\begin{equation}
\hat{c}(t) = \frac{c(t)}{a(t)} = \frac{c_0 S(t)}{a(t)}
\label{cComPhys}
\end{equation}
and the fact that $S(t)$ is a non-growing function.

For under-Brownian (as well as for Brownian and marginally over-Brownian) expansions we know that $\tau(t)\to \infty$ for $t\to \infty$ and that, in this case,  $c\sim  \tau^{-1/2}$.  On the other hand, for over-Brownian expansions we know that the reactions  eventually stop and thus $c=c_\infty>0$ for $t\to \infty$.  This means that in physical space one has $\hat c\equiv \hat c_\text{uB}\sim  1/[a(t) \tau^{1/2}]$ for under-Brownian expansions and $\hat c\equiv \hat c_\text{oB}= c_\infty/a(t)$ for over-Brownian expansions. In the latter case, the concentration decay is dominated by the dilution effect arising from the domain growth. This is easy to understand:  over-Brownian expansions are so fast,  that eventually the diffusion process becomes largely irrelevant, since the particle separation is mainly given by the deterministic domain growth, and the stochastic contribution arising from particle diffusion only represents a minor correction.

In particular, from Eqs.~\eqref{powerlawconftime} and \eqref{tauExp} one finds the following long-time behavior
\begin{equation}
\hat c_\text{uB} \sim
 \begin{cases}
 t^{-1/2} ,\quad \gamma<1/2,\\
 1,\quad H<0,\\
 \end{cases}
\end{equation}
(in fact, $\hat c_\text{uB} \approx 1/\sqrt{-2H}$ when $H<0$)
for power-law and exponential expansions, respectively.
In the opposite case of over-Brownian expansions one obtains
\begin{equation}
\hat c_\text{oB} \sim
 \begin{cases}
 t^{-\gamma} ,\quad \gamma>1/2,\\
 \exp(-Ht),\quad H>0.\\
 \end{cases}
\end{equation}
Note that all the results stated in this subsection for under-Brownian expansions also hold for the Brownian and marginally over-Brownian ones. At this stage, a remark on the validity of the mean-field approximation for the case of a power-law expansion is in order. In Ref.~\cite{OurPreprint2018}, it has been recently shown for the annihilation reaction that a mean-field-like rate equation accounting for the effect of a power-law domain growth (but neglecting the effect of fluctuations) yields $\hat c \propto t^{-1}$ for $\gamma<1$ and $\hat c \propto t^{-\gamma}$ for $\gamma>1$. For the coalescence reaction, the same mean-field behavior is obtained. From the above expressions, one can see that agreement with the mean-field prediction is obtained for a sufficiently fast expansion, i.e., for $\gamma>1$. This is clearly different from the static case, for which the crossover dimension for mean-field behavior is $d_c=2$.

Finally, for a near power-law domain growth, we also give some results in the neighborhood of the Brownian case $\gamma=1/2$ for the sake of completeness. As an example, let us consider the behavior of the conformal time in the following three cases:
\begin{equation}
\tau(t) \sim
 \begin{cases}
 [\ln(t)]^2 ,\quad a(t) \sim [t/ \ln(t)]^{1/2},\\
 \ln(t), \quad a(t) \sim t^{1/2}, \\
 \ln[\ln(t)],\quad a(t) \sim [t \ln(t)]^{1/2},\\
 \end{cases}
\end{equation}
respectively leading to the following long-time behavior of the physical concentration:
\begin{equation}
\hat c_\text{m} \sim
 \begin{cases}
 [t \ln(t)]^{-1/2} ,\quad a(t) \sim [t/ \ln(t)]^{1/2},\\
 [t \ln(t)]^{-1/2}, \quad a(t) \sim t^{1/2}, \\
 [t \ln(t) \ln[\ln(t)] ]^{-1/2}, \quad a(t) \sim [t \ln(t)]^{1/2}.\\
 \end{cases}
\end{equation}
This behavior shows the nontrivial effects of different types of expansion in the vicinity of the Brownian case $a(t) \propto t^{1/2}$. In particular, the first case in
the above equation corresponds to an under-Brownian
expansion. Despite the fact that such an expansion is slower than a Brownian expansion (the second case in the above equation), the time decay of the concentration is the same in both cases. Finally, the third case
(marginally over-Brownian expansion) does not only correspond to a faster expansion, but also to a faster concentration decay. Nevertheless, inasmuch as one also has $\tau_\infty=\infty$ in this case, this type of expansion is qualitatively similar to the two previous cases (see Table \ref{tablaExptypes}). As we have seen, minute deviations from this law may induce a rich phenomenology in our system.

Let us now analyze in more detail the case of an exponentially contracting domain, $a(t)=\exp(Ht)$ with $H<0$. This case is characterized by the fact that the dilution effect induced by the reaction is counterbalanced by the contraction of the domain, in the sense that the physical concentration tends to a constant value at long times. Indeed, for large $t$ one has
\begin{equation}
\tau(t) = \frac{1- \exp(-2 H t)}{2 H} \approx -\frac{\exp(-2 H t)}{2 H},
\end{equation}
and consequently, from Eq.~\eqref{ctAsin},
\begin{equation}
c(t) \approx \frac{1}{\sqrt{2 \pi D\tau}} \approx \sqrt{\frac{-H}{\pi D}} \, \exp(Ht).
\end{equation}
This implies that the concentration in physical space  $\hat c(t)=c(t) \exp(-H t)$ goes to a finite value of order $1/\ell=\sqrt{-H/D}$. This result can be understood as follows. To begin with, we note that $\ell$ is the contractive Hubble length introduced in Ref.~\cite{Yuste2016} for exponentially contracting domains (note the missing square root in the definition of $\ell$ as a result of a misprint in the aforementioned reference). This length characterizes the linear size of the region (or ``Brownian Hubble sphere'' in $1d$) around a random walker that she can hardly attain, no matter for how long she has been evolving. Therefore, $\hat c(t) \sim 1/ \ell$ at long times. In this time regime, the domain contraction in physical space implies that the latter is roughly covered by a tessellation of Hubble regions of size $\ell$.
When two particles approach each other as a result of the contracting drift associated with the Hubble flow, they coalesce according to the reaction A+A$\to$ A; therefore, their two corresponding Hubble regions coalesce too. The net result is a new arrangement consisting of a smaller number of Hubble regions tessellating the whole physical space, which in turn implies a constant concentration $\hat c$ of order $1/\ell$.

\subsection{Scaling analysis}

It is instructive to rederive the temporal dependence of the concentration given by Eq.~\eqref{ctAsin} by means of scaling arguments. As in the case of a static domain, such arguments are based on the volume swept by the diffusing particles up to a given time, which gives a notion of the number of particles that each of them has killed. However, in the case of a growing domain it is convenient to consider the motion of the particles in comoving space rather in physical space.

The linear size of the typical displacement of a diffusing particle in comoving space is  $\bar x(t)\approx \sqrt{D \tau(t)}$ \cite{Yuste2016}, and so the particle explores compactly a region whose size is of the order $\bar x(t)$. After a time $t$, all the particles in that region up to a single one are destroyed, resulting in the following concentration decay:
\begin{equation}
\label{ctEscala}
c(t)\sim 1/\bar x(t) \approx 1/\sqrt{D\tau(t)}.
\end{equation}
This agrees with Eq.~\eqref{ctAsin}. In the static case one has $\tau=t$, and the above argument reproduces correctly the known behavior $c(t)\sim t^{-1/2}$.   For over-Brownian expansions one has $c_\infty\approx 1/\sqrt{D\tau_\infty}>0$.  Note that similar scaling arguments can be directly applied to the case of an encounter-controlled annihilation reaction \cite{Ben-Avraham2005}. Hence, the above results can be straightforwardly extrapolated to the case of annihilation.  This results in a similar long-time dependence of the concentration, subsequently confirmed by asymptotically exact calculations and numerical simulations (see Sec.~\ref{secAniqui}).

\subsection{Spatial ordering: behavior of the IPDF}

Let us now discuss the influence of the expansion process on the spatial ordering of the particles. As is well known, the coalescence reaction on a static domain is an example of dynamic self-ordering \cite{Ben-Avraham2005}, implying that a certain degree of spatial order is induced by the reactions in a system where the initial distribution of particles is fully disordered (Poisson distribution). This initial condition is mathematically described by a purely exponential IPDF $p(x,0)=c_0 e^{-c_0 x}$. Its non-zero value for sufficiently small $x$ reflects the existence of  particle clustering. Such clusters are quickly reduced as a result of reactions between nearby particles, thus giving rise on short length scales to inter-particle gaps that slow down the reaction with respect to the prediction of the law of mass action \cite{Sancho2007}. In the course of time, the typical interparticle distance grows due to the two mechanisms at play, reaction and domain growth, the former being influenced by the latter. Reactions between neighboring particles give rise to gaps that are subsequently enlarged by the expansion, which in turns reduces the reactivity of the system. What we have found in Sec.~\ref{secCoaguGF} is that the expression describing this behavior can be obtained from the IPDF for the static case [see Eq.~\eqref{pxtPoisson}] via the double transformation $y\to x$ and $t\to \tau(t)$, i.e., by a convenient contraction of the domain and a slowing down of the time flow. As for the spatial distribution of particles described by the IPDF (and also by $n$-point correlation functions, see Sec.~\ref{secCorrelation}), it is worth noting that the expansion only increases the values of the interparticle distance, but does not change the proportion of large gaps with respect to small ones.  Therefore, in terms of the static representation provided by the comoving coordinates, the gap distribution at a given Brownian conformal time is the same as the one obtained for a non-growing domain at time $t$.

\begin{figure}[t]
\begin{center}
        \includegraphics[width=\ancho\textwidth,angle=0]{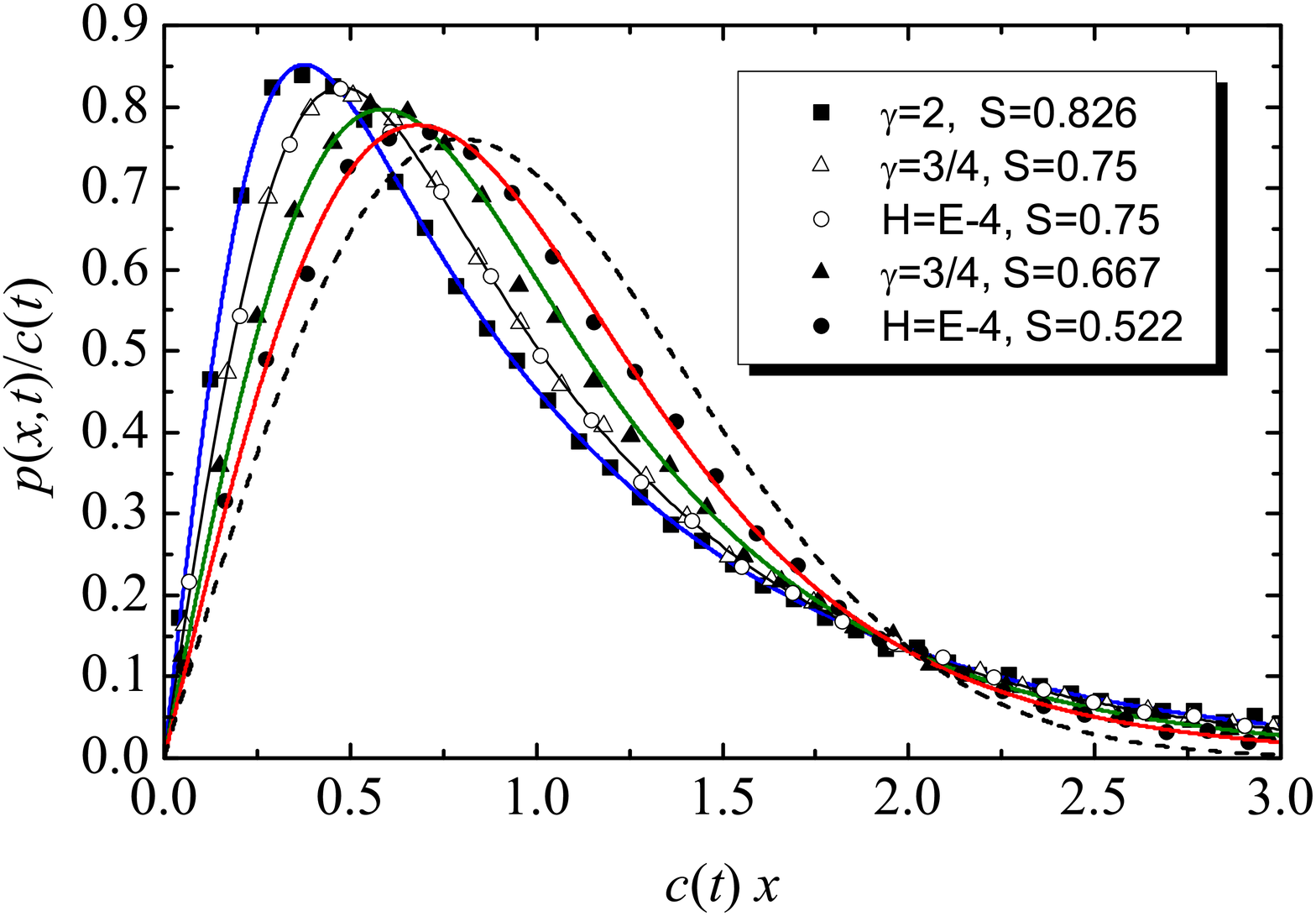}
\end{center}
\caption{\label{Figpxt2Coagu}
Scaled IPDF  vs.~$c(t)x$ for coalescence and three different over-Brownian expansions.   The thick lines from left to right near the origin depict the limiting IPDFs respectively obtained for $t\to\infty$ in three different cases of power-law expansion, i.e., $\gamma=2$ ($S_\infty\approx 0.823$), $\gamma=3/4$ ($S_\infty\approx 0.645$), as well as for an exponential expansion with $H=10^{-4}$ ($S_\infty\approx0.523$). The thin line is the IPDF for a time at which $S(t)=0.75$ whereas the dashed line is the long-time asymptotic IPDF for the static case. The symbols are simulation results for (i) power-law expansions with  $\gamma=2$ and a time $t$ such that $S(t)=0.826$, $\gamma=3/4$ and $S(t)=0.75$,  $\gamma=3/4$ and $S(t)=0.667$ (with $t_0=1000$ in all cases), and (ii) exponential expansions where $H=10^{-4}$ and $S(t)=0.75$, and $H=10^{-4}$ and $S(t)=0.522$.
 }
\end{figure}

Figure~\ref{Figpxt2Coagu} shows the behavior of the IPDF in the over-Brownian case. In addition to the significant decrease in the number of reactions and the associated non-empty steady state caused by the domain growth, the dynamic self-ordering process stops before the asymptotic scaling form given by Eq.~\eqref{pxtAsin} can be reached (the latter is represented by a thick solid line in the figure).  In other words,  the over-Brownian expansion frustrates the self-ordering process, and the system freezes in a state of intermediate ordering.  This behavior is reminiscent of the freeze-out events occurring in the fast expanding early universe \cite{Rehm1998,Ryden2003,Perkins2008}. As a result of the freeze-out, general expressions from which the behavior for small and intermediate values of $\tau$ can be inferred attain a somewhat unexpected relevance. Indeed, even if one is only interested in the long-time behavior of the reaction kinetics and the associated spatial distribution of particles, depending on the parameter choice, $\tau$ may freeze at small or moderate values, implying that $\tau_\infty$ is restricted to not too large values. When this happens, the concentration and the IPDF also freeze. However, since $t$ is very large, but $\tau$ is not, these quantities must be computed by replacing $t$ with $\tau$ in the general formulae for the static case, not in the long-time asymptotic formulae. A consequence of the above is that the self-ordering of the system is frustrated and the well-known long-time asymptotic scaling form $p(z,t\to \infty)\approx \pi (z/2) \exp\{-\pi (z/2)^2 \}$ with $z=c(t)x$ describing a highly ordered state is never attained. Simulations confirm the analytical results for the IPDF.

\begin{figure}[t]
\begin{center}
        \includegraphics[width=\ancho\textwidth,angle=0]{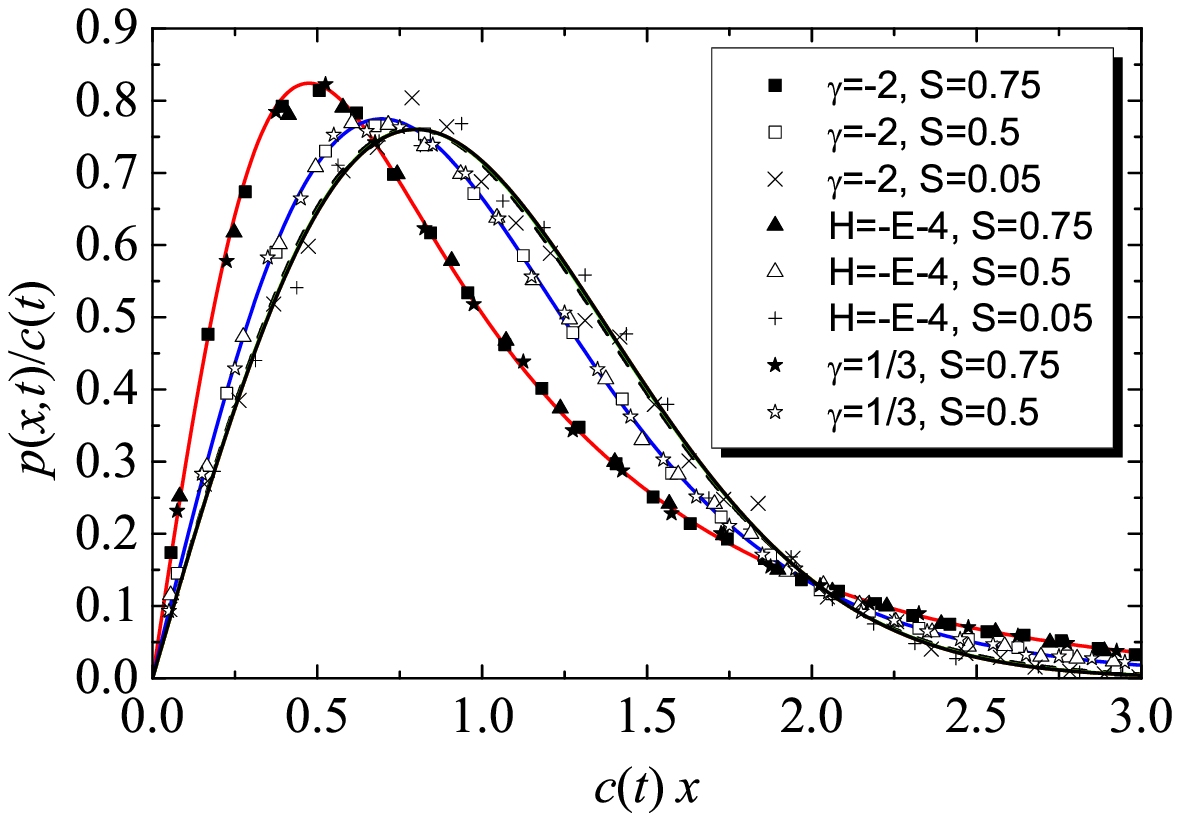}
\end{center}
\caption{\label{Figpxt1Coagu}
Scaled IPDF  vs.~$c(t)x$ for coalescence and three different under-Brownian expansions.  The lines from left to right near the origin depict IPDFs for times at which $S(t)=\left\{0.75, 0.5, 0.25, 0.15, 0.05\right\}$ and, finally, the long-time asymptotic IPDF for the static case (dashed line). The four last curves are almost indistinguishable.
The symbols represent simulation results for (i) power-law expansions with  $\gamma=-2$ and $S(t)=\left\{0.75, 0.5, 0.05\right\}$, as well as $\gamma=1/3$ and $S(t)=\left\{0.75, 0.5 \right\}$  (in all cases we take $t_0=1000$)  (ii) an exponential expansion where $H=-10^{-4}$ and $S(t)=\left\{0.75, 0.5, 0.05\right\}$.
}
\end{figure}

In Fig.~\ref{Figpxt1Coagu} the behavior of the IPDF is shown for the case of under-Brownian expansion, implying that $\tau\to\infty$ as $t\to \infty$. Note that this includes both the case of a contracting domain and the static case (not shown). In these cases, the self-ordering process induced by the reactions is accelerated and the memory of the initial condition is lost more quickly, resulting in a stronger departure of the asymptotic distribution from the initial exponential distribution. While the qualitative behavior is similar to the one observed for a fast expansion, in the present case of slow expansion the encounter-controlled reactions do not stop until the system becomes empty. We remind the reader that the Brownian and marginally over-Brownian cases
reproduce the under-Brownian behavior.

\subsection{Multiple-point density correlation}
\label{secCorrelation}

So far, we have restricted our description to a rather specific subset of properties, namely,  $\hat c(t)$, $\hat p(y,t)$, and $\hat E(y,t)$ [or, equivalently, $c(t)$, $p(x,t)$, and $E(x,t)$]. A more detailed characterization of the spatial ordering induced by the reaction is given by the quantity $\hat E_n(y_1,\bar y_1,y_2,\bar y_2,\ldots, y_n,\bar y_n,t)$, which describes the probability that the intervals $[y_i,\bar y_i]$ are all simultaneously empty at time $t$ (the intervals are assumed to be nonoverlapping and ordered, i.e., $y_1<\bar y_1<\cdots y_n<\bar y_n$) \cite{Ben-Avraham1998c}. Knowledge of $\hat E_n$ allows one to obtain the $n$-point correlation function $\hat \rho_n(y_1,y_2,\ldots, y_n,t)$, that is, the density corresponding to the probability  of simultaneously finding $n$ particles at positions $y_1,y_2,\ldots y_n$. One has \cite{Ben-Avraham1998c}
\begin{equation}
\label{hatrhon}
\hat \rho_n(y_1,y_2,\ldots, y_n,t)=(-1)^n \left. \frac{\partial^n}{\partial y_1\cdots\partial y_n} \hat E_n(y_1,\bar y_1,y_2,\bar y_2,\ldots, y_n,\bar y_n,t)\right|_{y_1=\bar y_1,\ldots,y_n=\bar y_n}.
\end{equation}
In certain cases, e.g., for an initial Poissonian distribution,  one can compute $\hat E_n$ from $\hat E_1$ \cite{Ben-Avraham1998c}:
\begin{equation}
\hat E_2(y_1,\bar y_1,y_2,\bar y_2,t)=
 \hat E_1(y_1,\bar y_1,t)\hat E_1( y_2,\bar y_2,t)
-\hat E_1(y_1, y_2,t)\hat E_1(\bar y_1,\bar y_2,t)
+\hat E_1(y_1, \bar y_2,t)\hat E_1(\bar y_1,y_2,t).
\end{equation}
The explicit expression for $\hat E_2(y_1,\bar y_1,y_2,\bar y_2,t)$ can then be obtained from the explicit formula for  $\hat E_1(y,\bar y,t)\equiv \hat E(\bar y-y,t)$   [see Eq.~\eqref{Extau}], and hence, from Eq.~\eqref{hatrhon}, one finds the two-point correlation function:
\begin{equation}
\label{hatrho2t}
\frac{\hat \rho_2(y_1,y_2,t)}{c_0^2}= \frac{1}{2} \text{erfc}\left(\frac{y_2-y_1}{\sqrt{8Dt}}\right) (Q_{-1}-Q_{1})+Q^2_{0}-Q_{1}Q_{-1},
\end{equation}
where $Q_{\alpha}\equiv Q_{\alpha}(y_2 - y_1 ,t)$ and
\begin{equation}
Q_{\alpha}(y,t)=
\exp\left(\alpha c_0 y + 2c_0^2Dt\right)
\text{erfc}\left(\frac{\alpha y + 4c_0 D t}{\sqrt{8Dt}}\right) .
\end{equation}
From this expression, Eq.~\eqref{Stazar}, and Eq.~\eqref{cComPhys}, one can straightforwardly find an explicit expression for the pair correlation function:
\begin{equation}
\label{hatgt}
\hat g(y_1,y_2,t)=\frac{\hat \rho_2(y_1,y_2,t)}{\hat \rho_2(y_1,y_1+z,t)_{z\to\infty}}=\frac{\hat \rho_2(y_1,y_2,t)}{\hat c(t)^2}.
\end{equation}
In the last equation we have used the fact that, for a completely random initial distribution of particles, one has $\hat{\rho}_2(y,y+\infty,t)=\hat{\rho}_1^2(y,t)=\hat{c}(t)^2$.

\begin{figure}[t]
\begin{center}
        \includegraphics[width=\ancho\textwidth,angle=0]{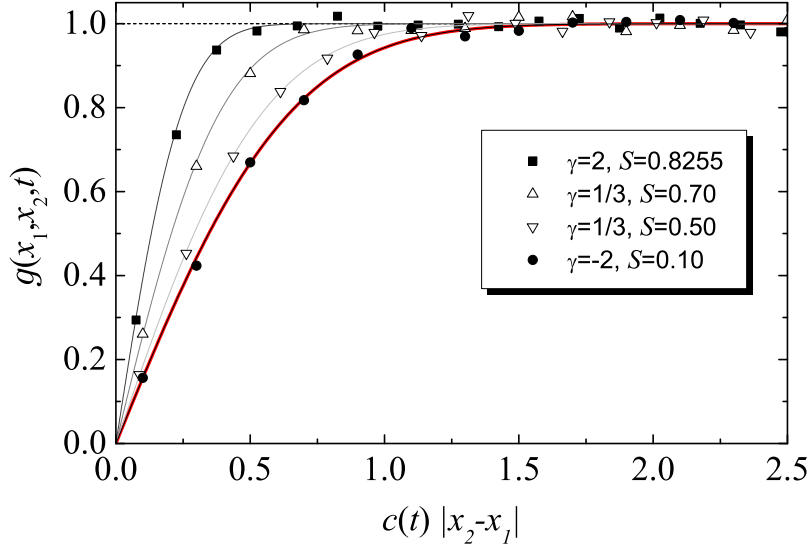}
\end{center}
\caption{\label{FiggxCoagu}
Two-particle correlation function $g(x_1,x_2,t)$ vs. $c(t)|x_2-x_1|$ for the coalescence process with an initial Poisson distribution of particles. The lines represent theoretical results [Eq.~\eqref{hatgt} with $y\to x$ and $t\to\tau(t)$] for power-law expansions with $\gamma=\{2,1/3,1/3,-2\}$ (from left to right near the origin). The times $t$ corresponding to these lines are those for which $S(t)=\{0.8255, 0.7, 0.5, 0.1\}$, respectively.   Symbols are simulation results corresponding to the aforementioned sets of values.  The line corresponding to $S(t)=0.10$ overlaps with the thick line representing the theoretical asymptotic pair correlation function obtained in the long time limit. The horizontal dashed line represents the pair correlation function for the initial Poisson distribution.
}
\end{figure}

Once again, the expressions for $E_n$, $\rho_n$, and $g$ in the case of an expanding domain are obtained by performing the change of variables $\{y\to x,t\to \tau\}$ in the corresponding expressions $\hat E_n$,  $\hat \rho_n$, and $\hat{g}$ for the static case.  For example, if $\tau(t)\to \infty$ when $t\to \infty$ one can find that the asymptotic pair correlation function  is given by \cite{Ben-Avraham1998c}
\begin{equation}
\label{rho2}
g(x_1,x_2,\tau\to\infty) \equiv g(x_1,x_2)= 1-e^{-2\xi^2}+\sqrt{\pi} \xi e^{-\xi^2} \text{erfc}(\xi)
\end{equation}
and  $\xi=(x_2-x_1)/\sqrt{8D\tau}$. Note that, for long times,  ${c}(t) \approx 1/ \sqrt{2 \pi D \tau}$   and, hence,   $\xi = (\sqrt{\pi}/2){c}(t)(x_2 - x_1)$.  In Fig.~\ref{FiggxCoagu} we show the excellent agreement between these analytical results and numerical simulations.

In order to understand the behavior shown in Fig.~\ref{FiggxCoagu}, let us first recall the behavior for the static case. For a random initial condition, there are no spatial correlations, and consequently one has $g(x_1,x_2,t=0)=1$. As the reaction proceeds, spatial correlations build up, since the dynamics of two particles after a given time is not independent. Indeed, two randomly chosen particles have a mutual influence on the survival probability of each other; at earlier times one of the chosen particles may have destroyed other particles that, were they to still exist, could have killed the other chosen particle. Clearly, this mutual influence decreases strongly with increasing distance, and $g$ tends to the unit function as $x_2-x_1$ becomes large. However, the region of rescaled distances for which $g$ becomes significantly smaller than one increases as the reaction proceeds and the concentration goes down. In fact, the correlation length is of the order of $1/c(t)$.

Let us now discuss the effect of the domain growth. For sufficiently long times, if the expansion is not fast enough to stop the reactions (i.e., for the under-Brownian case), the correlation function tends to the asymptotic form \eqref{rho2} (corresponding to the thick line in Fig.~\ref{FiggxCoagu}). In the opposite case of a sufficiently fast expansion (over-Brownian case), a few reactions take place at short times, but the number of reactions at longer times becomes negligible, resulting in a nonvanishing value of $S_\infty$. Consequently, the self-ordering process does not fully develop, implying that the long-time asymptotic spatial structure differs significantly from the prediction given by Eq.~\eqref{rho2}, and the system eventually adopts a frozen spatial structure corresponding to the limiting value $c_\infty=c_0 S_\infty$ of the concentration. This is e.g. the case for $\gamma=2$, which corresponds to the top curve and to the solid squares in Fig.~\ref{FiggxCoagu}. This curve and the squares correspond to a time for which  $c(t)/c_0= 0.8255$, which is very close to the final value $S_\infty \approx 0.823$. Again, the Brownian and marginally over-Brownian cases qualitatively behave as the under-Brownian case.

\section{Annihilation reaction on a uniformly growing domain}
\label{secAniqui}

For the diffusion-limited annihilation reaction on a static domain, Masser and ben-Avraham \cite{Masser2000,Masser2001,Masser2001a} developed a method of analysis very similar to the empty interval method, namely, the so-called method of odd/even intervals. Here, the function analogous to $\hat E(y,t)$ in the IPDF method is $\hat G(y,t)$, that is, the probability that an arbitrarily chosen segment of length $y$ contains an even number of particles.  By using arguments very similar to those employed for the coalescence reaction, Masser and ben-Avraham showed that $\hat G(y,t)$ obeys the same equation as $\hat E(y,t)$, but subject to a different boundary condition, i.e., $\hat G(0,t)=1$ and $\hat G(\infty,t)=1/2$.  The concentration $\hat c$ follows from the relation $\hat c(t)=-\partial \hat G(y,t)/\partial y|_{y=0}$, which is similar to Eq.~\eqref{cfromE}. For a completely random initial condition one has $\hat G(y,0)=1/2+\exp(-2 c_0 y)/2$,
and the survival probability $S(t)$ can be shown to be given by Eq.~\eqref{ctSt}, but in this case one takes $z=2c_0 \sqrt{2Dt}$. As shown in Ref.~\cite{OurPreprint2018} (where only the case of a power law expansion was considered), the replacement $t\to \tau$ allows one once again to obtain the exact solution.

\begin{figure}[t]
\begin{center}
        \includegraphics[width=\ancho\textwidth,angle=0]{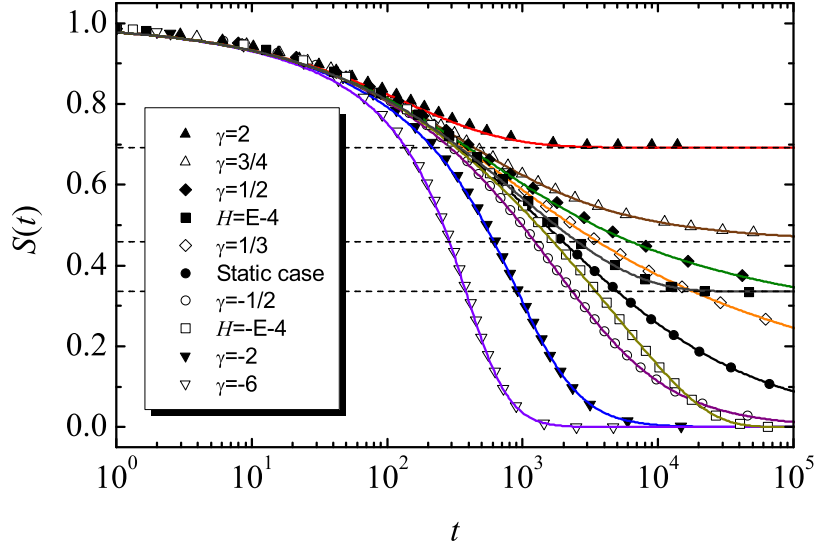}
\end{center}
\caption{\label{FigSttauLimAnni}
Time dependence of the survival probability for the annihilation reaction on a $1d$ domain subject to either an exponential expansion, to a power-law expansion with $t_0=1000$, or to a logistic expansion with $a_\infty=3$ and $\alpha=10^{-4}$. The symbols represent simulation results. Solid lines represent the corresponding theoretical predictions.  Horizontal dashed lines represent the asymptotic survival probability $S_\infty$ for the over-Brownian cases, namely, $S_\infty\approx 0.692$ for  $\gamma=2$,  $S_\infty\approx0.458$ for $\gamma=3/4$, and $S_\infty\approx0.336$ for  $H=10^4$.
}
\end{figure}

In Fig~\ref{FigSttauLimAnni}, this theoretical result for $S(t)$ is compared with simulation results for domains displaying power-law, exponential and logistic growth. In all cases, the agreement is once again excellent. Note that in the over-Brownian case, $S(t)$ converges to an asymptotic value that depends on the expansion rate, as is the case for the coalescence reaction. In the present case of annihilation, the decay to the steady state is faster than for coalescence, since particles always disappear pairwise. Note, however, that the asymptotic relation $\left.S(t)\right|_{coal.}\approx \left. 2S(t)\right|_{ann.}$ valid for the long time regime in the static case does no longer hold. In particular, for over-Brownian expansions this implies $\left.S_\infty\right|_{coal.}\not\approx \left. 2S_\infty\right|_{ann.}$.

\begin{figure}[t]
\begin{center}
        \includegraphics[width=\ancho\textwidth,angle=0]{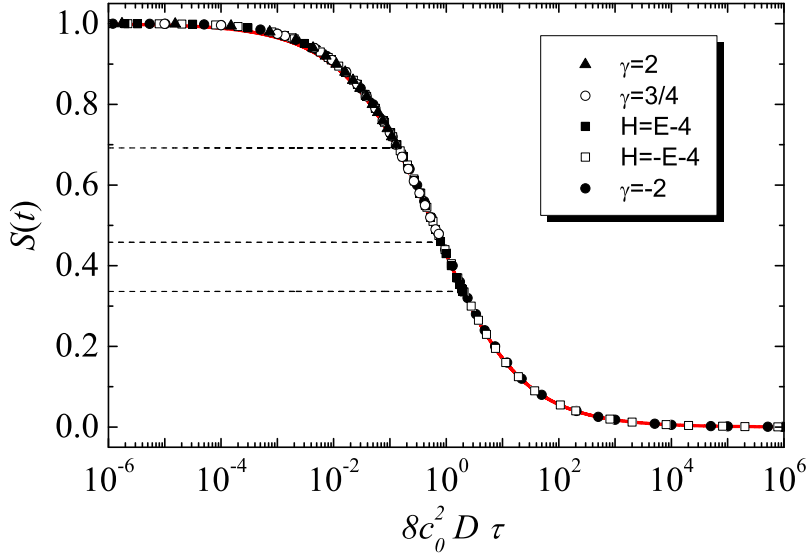}
\end{center}
\caption{\label{FigctAniqui}
Survival probability as a function of the scaling variable $z^2=8c_0^2 D \tau(t)$ in the case of annihilation on a growing domain.  The symbols represent simulation results.  We have used $t_0=1000$ for all the power-law expansions.
 Dashed lines represent the asymptotic survival probability $S_\infty$ for the over-Brownian cases ($\gamma=2$, $\gamma=3/4$ and $H=10^{-4}$).
 The solid line is the exact solution.
}
\end{figure}

Figure~\ref{FigctAniqui} shows the behavior of the survival probability when plotted vs.~the scaling parameter $z^2=8c_0 D\tau$. The universality of this curve must be interpreted in exactly the same way as for coalescence. The slowing-down of the reaction kinetics due to the domain growth is counterbalanced exactly by introducing a new time scale in which everything runs faster, i.e., a time scale defined by the $\tau(t)$.

Let us conclude the present section with some exact results concerning the space-time structure of the annihilation systems and valid for all times.
Here, the expressions $p(x,\tau)$, $G(x,\tau)$ and $ c(\tau)$ for the case of a growing domain are again obtained from their counterparts $\hat p(y,t)$, $\hat G(y,t)$ and $\hat c(t)$ in the static case by just performing the double replacement $y\to x$  and $t \to \tau(t)$.
 In the original work by Masser and ben-Avraham for the static case, no result for the IPDF was given. However, in a previous work, Alemany and ben-Avraham \cite{Alemany1995} had been able to obtain the long-time asymptotic form of the IPDF by mapping the annihilation reaction to the Glauber model. They showed how to relate the functions $\hat p$ and $\hat G$.  In the case of an expanding medium, the same relation between $p(x,\tau)$ and $G(x,\tau)$ holds, i.e.,

\begin{equation}
\label{pstgral}
\tilde{p}(s,\tau)=\frac{1- h(s,\tau)}{1+ h(s,\tau)}
\end{equation}
with
\begin{equation}
\label{hstau1}
h(s,\tau)=\frac{s}{c(\tau)} \left[ 1-s \tilde{G}(s,\tau)\right],
\end{equation}
where $\tilde{p}(s,\tau)=\int_0^\infty e^{-s x}  p(x,\tau) \, dx $ and $\tilde{G}(s,\tau)=\int_0^\infty e^{-s x} G(x,\tau) \, dx $ denote the Laplace transforms of $p$ and $G$ with respect to the spatial variable. For a completely random initial condition, the relevant boundary value problem for $G$ leads to (cf. Eqs.~(13) and (22) in Ref.~\cite{Masser2001a})
\begin{align}
G(x,\tau)&=
\frac{1}{2}+\frac{1}{2}\text{erfc}\left(\frac{x}{\sqrt{8D\tau}}\right)-\frac{1}{4}e^{8Dc_0^2 \tau} \nn \\
\label{gtau}
& \times \left\{ e^{2c_0 x} \left[1-\text{erf}\left(\frac{y+8Dc_0\tau}{\sqrt{8D\tau}}\right) \right]-e^{-2c_0 x} \left[1+\text{erf}\left(\frac{x-8Dc_0\tau}{\sqrt{8D\tau}}\right) \right]\right\}.
\end{align}
Hence, one obtains
\be
\label{hstau2}
 h(s,\tau)=\frac{s}{s^2-4c_0^2}\left[s-\frac{2c_0 e^{2D\tau(s^2-4c_0^2)} \, \text{erfc}(s\sqrt{2D\tau})}{\text{erfc}(2c_0 s \sqrt{2D\tau})}\right].
\ee
This equation, together with Eq.~\eqref{pstgral}, yields an exact expression for the IPDF valid at all times, albeit in Laplace space.  In the large-$\tau$ limit one obtains
\begin{equation}
\label{psAniqui}
\tilde{p}(s,\tau)\approx \frac{1-  s \sqrt{2\pi D\tau} \exp(2D\tau s^2)\, \text{erfc}(s\sqrt{2D\tau})}{1+s \sqrt{2\pi D\tau}   \exp(2D\tau s^2)\, \text{erfc}(s\sqrt{2D\tau})},
\end{equation}
which corresponds to a result first obtained by Alemany and ben-Avraham for the static case ($\tau=t$)~\cite{Alemany1995}. Note that, while Eqs.~\eqref{hstau1} and~\eqref{gtau} leading to Eq.~\eqref{hstau2} were already known from Refs.~\cite{Alemany1995} and~\cite{Masser2001a} in the static case,
we have not been able to find Eq.~\eqref{hstau2} in the literature. Note that Eq.~\eqref{hstau2} covers both the case of a static and expanding media over the whole time domain.

\begin{figure}[t]
\begin{center}
        \includegraphics[width=\ancho\textwidth,angle=0]{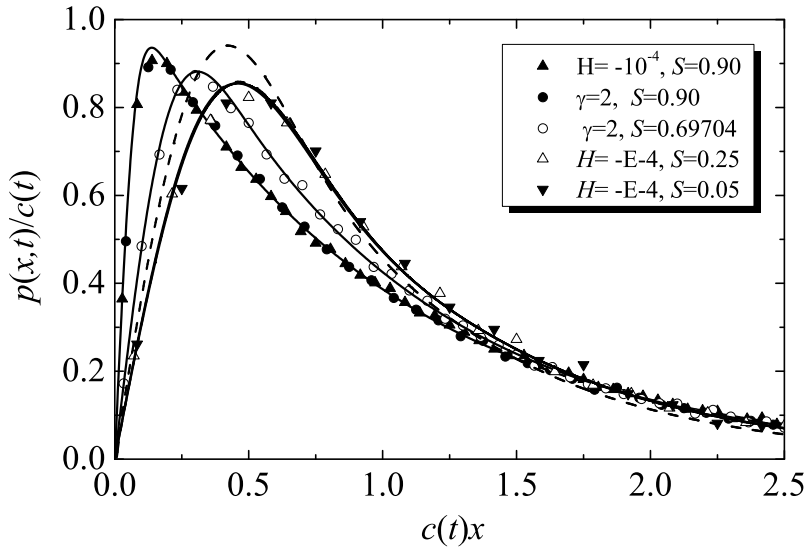}
\end{center}
\caption{\label{FigpxtAniqui}
Scaled IPDF vs.~$c(t)x$ for the annihilation reaction on a $1d$ growing domain. The symbols are simulation results for a power-law expansion with $\gamma=2$ and  $t_0=1000$ (squares),  as well as for an exponential contraction with $H=-10^4$ (triangles), at times where the survival probability $S$ is 0.9, 0.697, 0.25 and 0.05.  The solid lines are exact theoretical predictions for times at which $S$ is 0.9, 0.69195, 0.25 and 0.05 (from left to right near the origin). The two last curves are indistinguishable.  The value $S=0.69195$ is precisely the limiting value $S_\infty$ for $\gamma=2$.  The broken lines represent the analytical prediction~\eqref{psAniqui} at times for which $S=0.25$ and $S=0.05$. This last curve is indistinguishable from the asymptotically exact one.}
\end{figure}

In Fig.~\ref{FigpxtAniqui} these theoretical expressions for the IPDF are compared with simulation results for an over-Brownian power-law expansion and for an exponential contraction. The agreement is excellent. The qualitative behavior of the IPDF is essentially similar as in the case of the coalescence reaction, in the sense that in the over-Brownian case one also observes deviations from the long-time asymptotic scaling form obtained for the static case. The latter is characterized by a purely exponential tail $e^{-A c(t) x}$ \cite{Alemany1995}, where $A$ is a dimensionless constant whose value is well-known (recall that for coalescence one has a stretched exponential tail of the form $e^{-(\pi/2) c(t)^2 x^2/2}$). In the opposite case of an under-Brownian expansion, the numerical inversion of the asymptotic formula Eq.~\eqref{psAniqui} works fairly well for times at which $S(t)$ is about 0.05 or smaller.

\section{Conclusions and outlook}
\label{secConc}

The $1d$ kinetics of the coalescence and annihilation reactions (including the spatial organization of the reactants) have been the subject of many studies in the past. Different techniques have been developed, notably the IPDF method, which is among the most powerful ones. The IPDF method provides an exact solution both for the A+A$\to$ A and the A+A$\to \emptyset$ reaction.  In this paper, we generalize the IPDF method to study the impact of a uniform growth (contraction) process of the spatial domain on the reaction kinetics, both at the level of the particle concentration and of the spatial organization of the reactants. It should be noted that in Ref.~\cite{OurPreprint2018} the IPDF method was used to carry out a partial study focusing on the concentration behavior of the annihilation reaction in a $1d$ domain subject to a power-law expansion. Our findings in the present work confirm quantitatively the intuitive expectation that the observed slowing-down of the kinetics is due to changes in the spatial organization of the particles and, in particular, to the increasing separation between them.

The key to obtain the exact solution is a double transformation in time and space, whereby one switches from physical space and time variables ($y, t$) to a representation in terms of comoving coordinates and of the Brownian conformal time $(x,\tau)$. With this mapping $(y,t)\to (x,\tau)$, the IPDF equations take the same mathematical form as in the case of a static domain, and so the well-known solutions for the latter case can be used. However, some care must be taken when considering the long-time behavior in the case of a fast expansion (over-Brownian case). Here, the reactions stop before the system can become empty, and the particle concentration $c_\infty$ (measured in units of comoving length) tends to a finite value. If one conveniently scales out the length dilation introduced by the over-Brownian domain growth, one finds that the IPDF adopts a frozen profile corresponding to the case of a non-growing domain whose particle population is characterized by the concentration value $c_\infty$.  In this case, the self-organization process induced by the transport-mediated reactive events is frustrated at an early stage by the expansion, as a consequence of which this self-organization process does not fully develop. In the above context of an over-Brownian expansion, it is necessary to pay special attention to the equations describing the early- and intermediate time regime in the static case, which are usually not the main focus on the literature. However, we have seen that such equations are relevant to obtain the long-time kinetics on a sufficiently fast growing domain.

Summarizing, we have obtained explicit exact results for key quantities characterizing the $1d$ reaction kinetics of the coalescence/annihilation systems. For coalescence, we have computed exactly the particle concentration (both in comoving and physical space), the IPDF $p(x,t)$, the empty interval probability $E(x,t)$, as well as the pair correlation function $g(x,t)$, including the limits of these quantities when $t\to \infty$. As we have seen, the behavior of the concentration can also be obtained from simple scaling arguments which are expected to work in higher dimensions too.
Let us briefly recall the findings that emerge from the above analysis. First, we stress that the obtained results hold for arbitrary (uniform) expansion processes. We have considered the special cases of a power-law growth, of an exponential growth/ contraction and, to a lesser extent, the case of a logistic growth. For a sufficiently fast expansion (over-Brownian case), the behavior of the system changes dramatically with respect to the static case, e.g., a non-empty final state is attained; even though the time evolution and the final value of the survival probability are very sensitive to the details of the expansion, here we have seen that they can be computed from a universal formula valid for any continuous function $\tau(t)$. We have also discussed in detail the long-time behavior of the concentration in physical space. In the case of an exponentially contracting domain, we find the interesting result that the asymptotic concentration goes to a constant value, i.e., the particle loss due to the ongoing reactions is balanced by the strong decrease of the system's volume. Secondly, we have seen that the time evolution of the concentration/survival probability depends strongly on the details of the initial condition. In the vicinity of the marginal case separating over-Brownian from under-Brownian expansions, we have studied the transient behavior and we have seen that the relaxation time to reach the steady state may become extremely large. In all cases, we have performed simulations as well as a scaling analysis to confirm the above results.

The description of the spatiotemporal organization of the A+A systems is the main goal of the IPDF method, not only because the intrinsic interest of these quantities (the $n$-point correlation function or the IPDF itself, say),  but also because other relevant quantities such as the particle concentration depend on them [see, e.g., Eqs.~\eqref{comconc} and \eqref{Eyt}-\eqref{pfromE}]. For example, in some A+A systems the concentration behavior is very sensitive to the details of the initial particle distribution \cite{Ben-Avraham2005,Burschka1989,Ben-Avraham1990}. In this paper, we have also devoted special attention to the time evolution of the spatial distribution of particles in an expanding medium. Specifically, we have presented exact results for multiple-point density correlation functions. In particular, explicit exact results for the two-point correlation function in the case of an initial Poisson distribution of particles have been given. We find that the self-ordering properties of the system are strongly affected by the domain growth, implying that a certain degree of disorder may persist for very long times when the system starts from a completely random initial condition. As we have seen here, the asymptotic scaling form of the IPDF characteristic of the static case is not necessarily reached in the over-Brownian case.

For the annihilation reaction, we have obtained exact results for the concentration and the survival probability, including the final value of the latter. We have compared the behavior of the long-time survival probability in the case of the coalescence reaction with that of the annihilation reaction, and we have seen that the expansion process breaks the rule of thumb $S(t)|_{coal.}\approx 2S(t)|_{ann.}$ at long times. We have also obtained an expression for the Laplace-transformed IPDF, from which the behavior of the IPDF itself can be easily inferred.

In the above context, we distinguish between four regimes (over-Brownian, marginally over-Brownian, Brownian, and under-Brownian), which can be grouped in two different qualitative behaviors. In the over-Brownian case (fast expansion) explicit results for the freeze-out survival probability $S_\infty$, the IPDF $p(x,\tau\to\infty)$ and the pair correlation function $g(x,\tau\to\infty)$ have been derived. The over-Brownian case is the one exhibiting the strongest deviation from the behavior in the standard case of a static domain; in contrast, the qualitative behavior in the remaining three cases is similar. In all cases, the underlying physics has been comprehensively discussed.

With regard to possible extensions of the present work, an open question is whether the space-time transformation $(t\to\tau(t), y\to x)$ used here can be generalized to deal with other systems amenable to exact solution via the IPDF method, i.e., systems incorporating the back reaction A+A$ \rightleftarrows$A  \cite{Burschka1989} and/or particle birth  \cite{Doering1989}.
The latter two processes introduce additional time scales (associated with the inverse of the back reaction rate and of the birth rate) which may complicate the solution on a growing domain. In contrast, other variants such as finite systems \cite{Mandache2000}, or systems with inhomogeneities \cite{Doering1991} or traps \cite{Ben-Avraham1998b} are expected to be amenable to exact solution in the framework of the IPDF formalism.

Finally, when it comes to describing the kinetics of the coalescence/annihilation reactions on growing domains of arbitrary dimensionality, it is important to determine the domain of validity of mean-field approximations. As already mentioned in Ref.~\cite{OurPreprint2018}, one such approximation was developed by adding to the standard quadratic concentration decay term in the rate equation an additional term describing the system's dilution induced by the domain growth. As a result of this, the long-time decay of the system was seen to follow this mean-field approximation already in $d=1$ \cite{OurPreprint2018}. Scaling approaches are, on the other hand, expected to yield the correct time dependence of the concentration in higher dimensions. This will allow one to establish the precise parameter regime in which mean-field behavior is expected to hold, a central question in this field of research. Work in this direction is already underway.

\section{Acknowledgements}

This work was partially funded by the Spanish Agencia Estatal de Investigaci\'on through Grants No. MTM2015-72907-EXP (C.~E.) and FIS2016-76359-P (partially financed by FEDER funds) (S.~B.~Y. and E.~A.), as well as by the Junta de Extremadura through Grant No. GR18079 (S.~B.~Y. and E.~A.). F.~L.~V. acknowledges financial support from the Fundaci\'on Tatiana P\'erez de Guzm\'an El Bueno and from the Junta de Extremadura through Grant. No. PD16010 (FSE funds). C.~E. is grateful to the Departamento de F\'{\i}sica of the Universidad de Extremadura for its hospitality.

\end{document}